\documentclass[usenatbib]{mn2e}
\usepackage{times}
\usepackage{graphicx} 
\usepackage{amsmath}
\usepackage{mathabx}
\usepackage{subfigure}
\usepackage{natbib}
\usepackage{dblfloatfix}
\usepackage[T1]{fontenc}
\usepackage{hyperref}

\begin{document}

\title[GEEC2]{The GEEC2 spectroscopic survey of Galaxy Groups at $0.8<z<1$}
\author[Balogh et al.]{Michael L. Balogh$^{1,2}$, Sean L. McGee$^{2}$,  Angus Mok$^{1,3}$, David J. Wilman$^{4}$, 
\newauthor Alexis Finoguenov$^{5}$, Richard G. Bower$^{6}$, John~S. Mulchaey$^{7}$, Laura C. Parker$^{3}$
\newauthor Masayuki Tanaka$^{8}$
\\
$^{1}$Department of Physics and Astronomy, University of Waterloo, Waterloo, Ontario, N2L 3G1, Canada\\
$^{2}$Leiden Observatory, Leiden University, PO Box 9513, 2300 RA Leiden, The Netherlands\\
$^{3}$Department of Physics and Astronomy, McMaster University, Hamilton, Ontario, L8S 4M1 Canada\\
$^{4}$Max--Planck--Institut f{\" u}r extraterrestrische Physik, Giessenbachstrasse 85748 Garching Germany\\
$^{5}$Department of Physics, University of Helsinki, Gustaf H{\" a}llstr{\" o}min katu 2a, FI-00014 Helsinki, Finland\\
$^{6}$Department of Physics, University of Durham, Durham, UK, DH1 3LE\\
$^{7}$Observatories of the Carnegie Institution, 813 Santa Barbara Street, Pasadena, California, USA\\
$^{8}$National Astronomical Observatory of Japan, 2-21-1 Osawa, Mitaka, Tokyo 181-8588, Japan\\
}
\date{\today}
\maketitle
\begin{abstract}
We present the data release of the Gemini-South GMOS spectroscopy in the fields of 11 galaxy groups at $0.8<z<1$, within the COSMOS field.  This forms the basis of the {\it Galaxy Environment Evolution Collaboration 2} (GEEC2) project to study galaxy evolution in haloes with $M\sim 10^{13}M_\odot$ across cosmic time.  The final sample includes $162$ spectroscopically--confirmed members with $R<24.75$, and is $>50$ per cent complete for galaxies within the virial radius, and with stellar mass $M_{\rm star}>10^{10.3}M_\odot$.  Including galaxies with photometric redshifts we have an effective sample size of $\sim 400$ galaxies within the virial radii of these groups.  We present group velocity dispersions, dynamical and stellar masses.  Combining with the GCLASS sample of more massive clusters at the same redshift we find the total stellar mass is strongly correlated with the dynamical mass, with $\log{M_{200}}=1.20\left(\log{M_{\rm star}}-12\right)+14.07$.  This stellar fraction of $~\sim 1$ per cent is lower than predicted by some halo occupation distribution models, though the weak dependence on halo mass is in good agreement.  Most groups have an easily identifiable most massive galaxy (MMG) near the centre of the galaxy distribution, and we present the spectroscopic properties and surface brightness fits to these galaxies.  The total stellar mass distribution in the groups, excluding the MMG, compares well with an NFW profile with concentration $4$, for galaxies beyond $\sim 0.2R_{200}$.  This is more concentrated than the number density distribution, demonstrating that there is some mass segregation.  
\end{abstract} 

\section{Introduction}
It remains a challenge to understand how galaxies form and evolve within a $\Lambda$CDM Universe, as the rate of galaxy stellar mass assembly is observed to be largely decoupled from the gravitationally--driven structure formation rate of the underlying dark matter \citep[e.g.][]{Bower06,Behroozi13}.  In general terms, though, a fundamental prediction of $\Lambda$CDM is that structure grows hierarchically, and observations of galaxy clustering \citep[e.g.][]{Zehavi02,Tegmark04,Vipers_clustering,Kim14} confirm this.  
A successful description of this clustering is the halo model \citep[e.g.][]{Berlind02}, in which each virialized dark matter structure is assumed to host one galaxy at its centre, with a number of orbiting satellite galaxies that were accreted at some time in the past, and which will eventually merge with the central object.  This description works well, at least for massive galaxies $M>10^9M_\odot$; it may break down at lower masses where not all dark matter haloes host a galaxy \citep[e.g.][]{Sawala14}.

Observations of galaxies in groups and clusters have long demonstrated their power in unveiling some of the key processes of galaxy evolution.  They are systems in which the dark matter halo mass can be measured directly, through several independent methods.  Moreover the high virial masses means most of the baryons are visible, as the diffuse gas between galaxies is at densities and temperatures that makes it detectable by X--ray observatories.  The galaxy population itself provides a record of several Gyr worth of evolution prior to the epoch of observation, as the radial and velocity distribution of satellite galaxies depends on the epoch at which they were accreted \citep[e.g.][]{BNM,McGee09,Oman}.  The redshift dependence of these distributions can therefore probe the timescales associated with galaxy evolution \citep[e.g.][]{Ellingson,Wetzel14}.  Finally, the evolution in the space density of groups and clusters is a direct prediction of $\Lambda$CDM models, and can therefore provide tests of cosmological models \citep[e.g.][]{RBN,Voit_rev}.  Indeed, observations of galaxy clusters (both their abundance and their mass-to-light ratios) provided some of the first strong evidence for $\Omega_m<1$ \citep[e.g.][]{W93,CNOC}.

At low redshift, complete and comprehensive catalogues of galaxy groups and clusters spanning a wide range of halo masses have been constructed in many independent ways \citep[e.g.][]{REFLEX,JM03,OP04,2PIGG,REXCESS,YangGC}.  Particularly valuable have been wide-area, highly complete spectroscopic surveys like the 2dFGRS \citep{2dfgrs}, SDSS \citep{SDSS} and GAMA \citep{GAMA}.   The availability of spectroscopy for most galaxies first provides robust membership information, and good measurements of the dynamics and stellar content of dark matter halos of a given mass.  Moreover the spectra themselves reveal unique information about the galaxy stellar populations: their ages, metallicities and star formation rates \citep[e.g.][]{KE08,CGvD}.  

Cluster catalogues have also been compiled out to much higher redshift, based on detection in X--ray \citep[e.g.][]{RBN,MCXC}, SZ decrement \citep{SPT,SPT10,SPT11,Planck_cluster}, photometric overdensity \citep[e.g.][]{GY,MaxBCG,GMBCG,RCS2,madcows,Sparcs} or redshift-space clustering from spectroscopy \citep{CNOC2g,DEEP2g,Knobel12}.  In general the limiting detectable halo mass increases with redshift, while the spectroscopic coverage decreases due to sparse sampling of increasingly massive galaxies.  Nonetheless, follow-up observations of massive clusters is easily motivated, as their limited spatial extent and high overdensity makes spectroscopy very efficient.  Thus large spectroscopic samples of cluster galaxies exist for many clusters even beyond $z=1$ \citep[e.g.][]{Demarco07,Fassbender11,GCLASS,madcows14}, and these will continue to grow.

Analysis of distant massive clusters is valuable on many fronts.  But, in general, these systems are rare and not representative progenitors of the structures we observe today; clusters more massive than $\sim 5\times 10^{14}$ at $z=1$ are expected to be more massive than Coma by $z=0$.  To trace galaxy evolution through the hierarchical buildup predicted by $\Lambda$CDM \citep[e.g.][]{B+08} requires the study of {\it less} massive haloes at higher redshift, and this is considerably more challenging.  Identifying these systems in the first place is difficult, because their X--ray and SZ flux is much lower, their individual weak lensing signals are unmeasureable, and the lower stellar masses require highly complete and deep spectroscopy to pick them out of wide-field spectroscopic surveys.  Very deep X--ray observations do enable detection of low-mass clusters out to $z\sim 2$ \citep[e.g.][]{Fin07,Xray-aegis,Giodini,Tanaka13}, though the identification with optical overdensities can be challenging \citep{JC12}.  By stacking such systems it is possible to measure their mass via weak lensing \citep{COSMOS_lensing,GEEC1_MstarMhalo}.  Generally, however, groups are most efficiently selected from uniform redshift surveys \citep[e.g.][]{CNOC2g,Knobel09,VVDS,DEEP2g,Knobel12} or good photometric redshifts \citep[e.g.][]{PFOF,GH11,PFOF_zc}.

Once identified, follow-up spectroscopy of distant groups remains very difficult.  The low richness means spectroscopy must identify galaxies far down the mass function in order to return sufficiently large samples of galaxies to study individual systems.  But at these depths the images are dominated by galaxies in the fore- and background, and the efficiency of identifying cluster members is too low to be practical.  Considerable success has been attained through statistical analysis, by stacking many groups together and subtracting a statistical background population \citep[e.g.][]{Giodini,Knobel_qf,zC20_Kovac}.  Provided the purity and completeness of these samples are known, which requires comparison to realistic simulations, this can reveal the average characteristics of galaxy groups.  However, little if anything can be learned about their dynamics or the scatter from system to system \citep{MLB_scatter}; and the lack of spectroscopy limits what can be learned about the star formation histories or metallicities of the galaxies.

The Galaxy Environment Evolution Collaboration (GEEC) was formed to take on the task of obtaining large spectroscopic samples of group galaxies.  Our first endeavour used guaranteed time on the LDSS2 spectrograph on Magellan to observe $\sim 20$ groups at $0.3<z<0.5$, compensating for the the low efficiency of membership confirmation with a lot of telescope time \citep{GEEC1}.  The success of this project led us to a similar undertaking at $0.8<z<1$, dubbed GEEC2, and this is the subject of the present work.  The even lower efficiency expected at this redshift, and the longer integration times required, necessitated a change in strategy.  It is essential to use good photometric redshift preselection in order to improve the targeting efficiency.  Moreover, the telescope time is too valuable to risk on candidate groups that might prove to be chance projections of a few galaxies.  These requirements drove us to select groups from the COSMOS field, where deep X--ray data and sparse redshift coverage is available.  The initial observations were published in \citet[][Paper~I]{GEEC2}, and the first analyses of the stellar populations in these groups were published in \citet[][Paper~II]{Mok1} and \citet[][Paper~III]{Mok2}.  One of the main motivations for extending studies of environment to higher redshift is that the sensitivity to environment--driven galaxy transformations may be much higher, due to the faster average assembly rate (due to the shorter age of the Universe), the shorter dynamical timescale, and the greater star formation rates and gas richness of the field.  We found, in Paper~II, 
that the passive fraction in GEEC2 groups is very similar to that at low redshifts, a result that has been seen also in samples of groups with shallower, more sparsley sampled spectroscopy \citep{DEEP2_BO,Knobel_qf}, and in the more massive GCLASS clusters \citep{GCLASS}. Moreover, in Papers~I 
and II 
we showed that there exists an identificable population of intermediate--colour galaxies that lie between the red-sequence and blue cloud, and in Paper~III 
we use the abundance of this population to put strong constraints on the timescale for galaxy transformations.  

Here, in Paper~IV, we present a full description of the data, fundamental measurements of the integrated group properties, and the catalogue of sources with derived quantities.  
Throughout this paper we assume a WMAP9 \citep{WMAP9} cosmology ($H_\circ$=69.3km/s, $\Omega_m=0.286$, $\Lambda=0.713$), and projected distances are given in proper (physical) coordinates, not comoving.

\section{Data}
\subsection{Survey history and overview}
GEEC2 is a spectroscopic survey of galaxies in $11$ groups, one of which was serendipitously discovered in the background of the target, within the COSMOS field.  The spectroscopy was obtained with GMOS-South over two semesters (2010A and 2011A).  
The original goal of the survey was to observe $\sim 20$ groups, with $3$--$4$ spectroscopic masks each, to allow an investigation of the intrinsic scatter within group populations.  However, repeated attempts to complete the program have been thwarted by bad weather, scheduling conflicts at Gemini, and variance in ranking from semester to semester.  
Following the lack of any time awarded in 2012B, attempts to extend the sample have been abandoned for the moment.  

Details of the target selection and spectroscopic observations have been presented in Papers I--III.  We summarize most of the salient details, here.   
\subsection{Galaxy Group Selection}
Candidate group targets were selected from the COSMOS \citep{COSMOS} field, based on faint, extended sources detected from deep {\it Chandra} and {\it XMM} data \citep{Fin07}.  Specifically, we selected groups from a catalogue that was later published in \citet{George1,George2}.  Selection criteria were that they have redshifts $0.8<z<1$, confirmed with at least three spectroscopic members.  This initial spectroscopic identification was primarily from the 10K zCOSMOS survey \citep{zCOSMOS}, though other available spectra were also used.  Systems were only selected from the top two categories of robustness, representing secure detections with or without reliable X--ray centres.  Twenty-one systems fulfill these criteria.  As described above, sufficient time was only obtained to observe half the sample, and we prioritised the observations to focus on the lowest-mass, highest redshift groups in the list.  The only mass estimates available were from the X--ray luminosity, assuming the $z=1$ scaling relation of \citet{COSMOS_lensing}.  As we will show in \S~\ref{sec-smass}, these are in excellent agreement with the dynamical masses we measure from the GEEC2 spectroscopy.

\subsection{Spectroscopic Target Selection and Data reduction} \label{sec-specsel}
We use the existing deep photometric catalogues of \citet{Capak}, and select galaxies based on their {\it Subaru} $r+$ magnitude (hereafter just referred to as $r$), measured within a 3\arcsec\ aperture.  This choice was made deliberately to maximize the number of group members for which a redshift could be obtained.  Since $r$ corresponds to a blue rest-frame wavelength at $z\sim 0.9$, the sample is not mass-selected, but rather is dominated by low-mass, emission line galaxies.  This allows us to efficiently detect enough galaxies to measure a velocity dispersion, and a mass-limited sample can still be constructed from a subset of the data (see below).

The efficiency of GEEC2 spectroscopy is largely dependent on the exquisite photometric redshifts available, from 30 filters of various width \citep{Ilbert}.  The redshifts are estimated from a template--fiting technique, and have been calibrated from existing spectroscopy.   We use the published 68 per cent confidence limits ($\sigma_{\rm zphot}$) to select galaxies that have a photometric redshift within 2$\sigma_{\rm zphot}$ of the estimated group redshift.  Our highest priority targets are those with $21.5<r<24.75$. Secondary priority slits are allocated to galaxies with $15<r<24.75$ and $0.7<z_{\rm phot}<1.5$.  
At $z\sim 0.9$, the average uncertainty on $z_{\rm phot}$ increases from $\sigma_{zphot}\sim 0.007$ for the brightest galaxies to $\sigma_{zphoto}\sim 0.04$ for those at our limit of $r=24.75$. Even for these faintest objects, 90 per cent of the galaxies have $z_{\rm phot}$ uncertainties of less than $\sigma_{zphot}<0.07$.  In \citet{Mok1} we show that the spectroscopic redshifts we measure are generally in good agreement with the photometric estimates, and that few of the lower priority targets turn out to be group members.  Accounting for the sampling strategy, we estimated that $\sim 6$ per cent of the group population may be underrepresented in the spectroscopic sample, due to the photometric redshift preselection.  In this paper we take a different approach to correcting for incompleteness (see \S~\ref{sec-photoz}), which will mitigate this problem.

GMOS masks were designed using {\sc gmmps}.  Typically $50$--$70$ per cent of the $\sim 35$ available slits on the first mask are allocated to top priority targets.   This fraction decreases on subsequent masks, and in most cases three masks are sufficient to target at least $40$ of these galaxies.  Many masks for a given target use some of the same alignment stars and are always at the same position angle; thus a small fraction of the CCD area is unusable regardless of the number of masks obtained.  We use the nod-and-shuffle\citep{ns} technique with $1\times3$ arcsec slits, offsetting the target from the slit centre so it is observed in every frame.  The R600 grism and OG515 blocking filter were used for all observations.  However, the detector was binned in the spectral direction by $2$ and $4$ in 2010A and 2011A, respectively.  The resulting spectral resolution, limited by the slit width, is $\sim 12.8$\AA\ for the 2011A data, and $\sim 6.4$\AA\ for the 2010A data.    All masks were observed with two hours on-source exposure, in clear conditions with seeing $0.8$ arcsec or better.

All data are reduced in {\sc iraf}, using the {\sc gemini} packages with minor modifications as described in \citet{GEEC2}. Redshifts are measured by adapting the {\sc zspec} software, kindly provided by R. Yan, used by the DEEP2 redshift survey \citep{DEEP2}. This performs a cross-correlation on the 1D extracted spectra, using linear combinations of template spectra. The corresponding variance vectors are used to weight the cross-correlation. Finally, redshifts are adjusted to the local standard of rest using the {\sc iraf} task {\sc rvcorrect}, though this correction is negligible.

We adopt a simple, four-class method to quantify our redshift quality. Quality class 4 is assigned to galaxies with certain redshifts. Generally this is reserved for galaxies with multiple, robust features. Quality class 3 are also very reliable redshifts, and we expect most of them to be correct. These include galaxies with a good match to Ca H\&K for example, but no obvious corroborating feature.  The better spectral resolution of the 2010A data allowed us to resolve the [OII] doublet and thus obtain reliable redshifts based on this line alone.  For 2011A data, we assume that single emission lines are [OII], but assign a quality class 3.  Class 2 spectra are not considered reliable for scientific analysis and we do not consider them further.  Of the 810 unique galaxies for which a spectrum was successfully extracted, 603 have class 3 or 4 redshifts and are included in our analyses.  

\begin{table*}
\begin{tabular}{llllllllll}
Galaxy & RA       & Dec          & $z$ &  Quality&$\log(M_{\rm star}/M_\odot)$ &$W_\circ(OII)$&Class&Source\\
      &\multispan2{\hfil deg (J2000)\hfil} &           && &(\AA)&(p/sf/int)   \\
\hline
505476 & 150.39522 & 1.89073 & 0.9833 & 2.0 & $9.34^{+0.03}_{-0.06}$&$-$&sf&GEEC2\\ 
506821 & 150.45281 & 1.87985 & 0.2146 & 1.0 & $7.76^{+0.15}_{-0.28}$&$-$&sf&GEEC2\\ 
506963 & 150.43434 & 1.87900 & 1.0305 & 4.0 & $8.92^{+0.08}_{-0.13}$&$-$&sf&GEEC2\\ 
507468 & 150.43930 & 1.87472 & 0.8406 & 4.0 & $8.81^{+0.09}_{-0.09}$&$67.78\pm154.67$&sf&GEEC2\\ 
...\\
1969875 & 149.60096 & 2.80289 & 1.0147 & 9.5 & $10.26^{+0.08}_{-0.11}$&$-$&sf&zCOSMOS\\ 
1970058 & 149.70166 & 2.80139 & 0.3740 & 4.5 & $9.58^{+0.07}_{-0.10}$&$-$&sf&zCOSMOS\\ 
1970151 & 149.69347 & 2.80111 & 0.4904 & 2.5 & $10.22^{+0.07}_{-0.11}$&$-$&sf&zCOSMOS\\ 
\hline
\end{tabular}\caption{Sample entries from the online catalogue of spectroscopic redshifts are shown here. The full table includes the Galaxy ID, position on the sky, spectroscopic redshift and its quality flag, and the [OII] rest-frame equivalent width. The penultimate column provides our classification as passive (p), star-forming (sf) or intermediate (int), and the final column identifies the source of the redshift (GEEC2 or zCOSMOS).   While all GEEC2 data are included, we only include zCOSMOS spectra with good quality redshifts.  The full table contains 9245 entries.}
\label{tab-catalogue}
\end{table*}

Our sample also incorporates the DR2 release of the 10K zCOSMOS spectroscopic survey \citep{zCOSMOS}.  This provides redshifts for galaxies with $i<22.5$, with a $\sim 40$ per cent spectroscopic sampling completeness, for all groups. All galaxies with a high probability ($>90$ per cent) of being correct  are used.  Note that the redshift quality flags have a different definition from that used in GEEC2.
As reported in \citet{Mok1}, based on a few redundant observations we find the typical redshift uncertainty in GEEC2 is $\sim 100$ km/s, and there is a small, unexplained systematic offset compared with zCOSMOS, such that our redshifts are smaller by $\Delta z=6.2\times10^{-4}$.  As in \citet{Mok1}, we adjust the zCOSMOS redshifts by this amount so they are consistent with ours.

\subsection{Stellar Masses}\label{sec-smass}
Stellar masses for all galaxies with spectra were computed as described in \citet{Mok1}.  The mass--to--light ratios are obtained by fitting \citet{CB07} templates to all available photometry (FUV, NUV, U,B,V,G,R,I,Z,J,K and all four IRAC bands), following \citet{McGee11} and \citet{Salim}.  For the U through K bands we use the psf--matched photometry within 3\arcsec\ apertures, and apply an aperture correction given by the difference between $I_{\rm AUTO}$ (from on {\sc sextractor MAG\_AUTO}) and the aperture $I$ magnitude.  For the IRAC bands, the flux is computed within a 3.4\arcsec\ aperture, and corrected to total flux assume a point-like source, following \citet{Ilbert,Ilbert10}.  The corrections are 1.31, 1.35, 1.61 and 1.72 magnitudes, in order of increasing wavelength.  Finally, the catalogued FUV and NUV magnitudes from GALEX are total magnitudes, so no further correction was applied.  In any case, these wavelengths are irrelevant for the redshifts of interest here.

The templates include a range of star formation histories, including short bursts, and assume a universal \citet{Chabrier} initial mass function.

From the sample of galaxies with spectroscopic redshifts, we fit a bilinear relation to the correlation between $M/L$ in the IRAC [3.6]$\mu$m band (3.4\arcsec\ aperture magnitude) and the redshift and $(r-i)$ colour (bracketing the rest-frame 4000\AA\ break) of the galaxies:
\begin{equation}
\log{\left(\frac{M/M_\odot}{L/L_\odot}\right)}=0.61(r-i)-1.38z-0.62,
\end{equation}
where $L_\odot$ is the luminosity of the Sun at 3.6$\mu$m, adopting $6.061$ as the absolute magnitude ($L_{\odot,\nu}=1.675\times 10^{11}W/Hz$).  We use this to estimate the stellar mass of galaxies without a spectroscopic redshift.  The advantage of this approach is we can easily recompute the stellar mass for galaxies as they are probabalistically assigned to a given group (see below).

\subsection{Galaxy classification}
In \citet{Mok1} and \citet{Mok2} we described our procedure for classifying galaxies as star-forming, passive or intermediate, based on their colours.  It is now well-known that dusty, star--forming galaxies can be distinguished from truly passive galaxies in a colour-colour diagram, where an optical--IR colour is correlated with a colour bracketing the 4000\AA\ break \citep[e.g.][]{Labbe05,WGM,GEEC1_colours}.  We k-correct our colours to $z=0.9$ using the {\sc kcorrect} IDL software of \citet{kcorrect}, and use the resulting colours $(V-z)^{0.9}$ and $(J-[3.6])^{0.9}$ for our classification.  Star-forming galaxies are defined as those with $(V-z)^{0.9}<2$ or $(J-[3.6])^{0.9}>0.856\left[(V-z)^{0.9}-2.0\right]+1.008$; passive galaxies are those with $V(-Z)^{0.9}>2$ and $(J-[3.6])^{0.9}<0.856\left[(V-z)^{0.9}-2.0\right]+0.6311$, or $(V-z)^{0.9}>3$.  
There is a small colour range between these two populations, and we refer to galaxies in that space as intermediate type. 
As with the stellar masses (\S~\ref{sec-smass}), we fit these k-corrections as a function of $(r-i)$ and $z$ (redshift), though this time it is necessary also to include the cross-term $(r-i)z$.  This k-correction function is used to correct galaxies in the photometric redshift sample and separate them into the star-forming, passive or intermediate samples.

\subsection{Redshift catalogue}
The resulting catalogue of redshifts and key derived quantities is presented in Table~\ref{tab-catalogue}.  The galaxy position is taken from the photometric catalogue of \citet{Capak}.  We also provide the stellar masses, rest-frame equivalent width of the [OII]$\lambda$3727 emission line when present, and colour-based classification (as passive, star-forming or intermediate).  The Table includes galaxies from the zCOSMOS 10K data release with redshift quality flags indicating a $>90$ per cent probability of being correct; note that the redshifts have been adjusted as described in \S~\ref{sec-specsel}.  

\begin{figure*}
\centerline{\includegraphics[clip=true,trim=0mm 0mm 0mm 0mm,width=7in,angle=0]{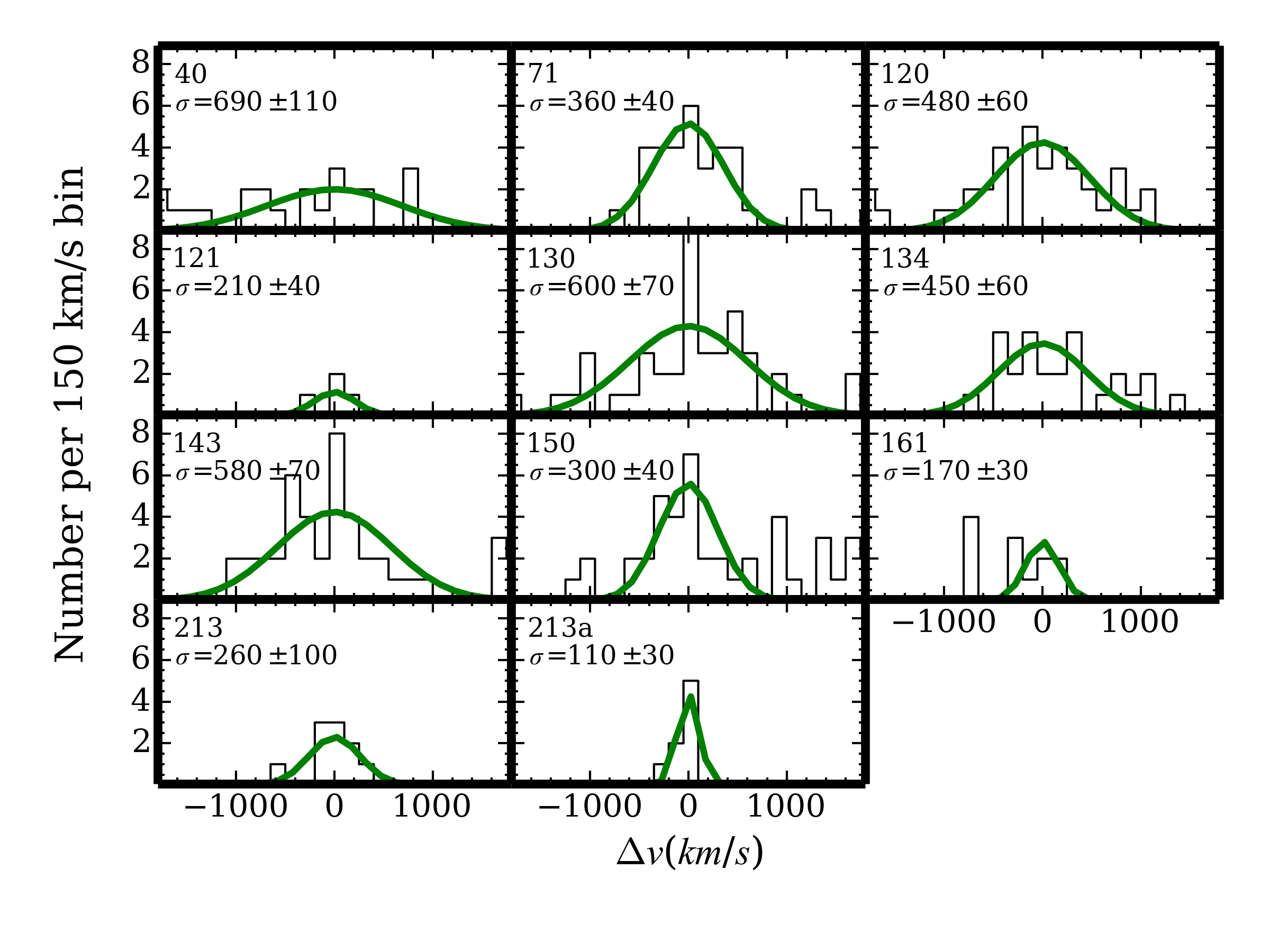}}
        \caption{Velocity distributions for the 11 groups, including all galaxies within $2R_{\it rms}$ of the mean galaxy position.  The group ID number, and its measured velocity dispersion (in units of km/s) are shown in each panel.  The green curve is a Gaussian function with this dispersion, normalized to match the total number of galaxies within $3\sigma$.  These are not the intrinsic velocity dispersions given in Table~\ref{tab-groups}, which are debiased and are the ones used to calculate dynamical properties.  Note that groups 150 and 161 are close together in space and velocity, with group 161 at the higher redshift.  In each system, the velocity peak from the neighbouring group is visible.}
\label{fig:vhist}
\end{figure*}

\section{Dynamics and Group Masses}
\subsection{Velocity Distributions}\label{sec-velocities}
For each group, we begin with the X--ray centre and the redshift estimated from the COSMOS group catalogue \citep{George1}.  Starting with all galaxies within $0.5$ Mpc with good quality redshifts, we measure the mean redshift $z_{\rm group}$ and position, the {\it rms} offset from the mean position, $R_{\it rms}$, and the velocity dispersion, using the gapper technique \citep{Beers}.  Preliminary group members are then identified as those within $2\sigma$ of the mean redshift, and within $2R_{\it rms}$ in position, and the calculation is repeated.  This process is repeated five times to determine the final $\sigma$ and $R_{\it rms}$.  For a few systems, the clipped values are modified.  Groups 150 and 161 are close together in space and redshift, and while their velocity distributions are clearly distinct, the clipping had to be stricter to exclude the other structure.  Group 213a has few members and a slightly larger clipping was adopted.

In Figure~\ref{fig:vhist} we show the rest-frame velocity distributions of each group.  All galaxies within 2$R_{\it rms}$ and secure redshifts are included, and a Gaussian with the measured $\sigma$ is overplotted for comparison.  In most cases the Gaussian is a reasonable fit; groups are dynamically well separated from their surroundings, and there is little subjectivity required to define the velocity limits for membership. Group members in the final analysis are taken to be those within $3\sigma$ of the median redshift.

\begin{figure*}
\centerline{\includegraphics[clip=true,trim=0mm 0mm 0mm 0mm,width=3in,angle=0]{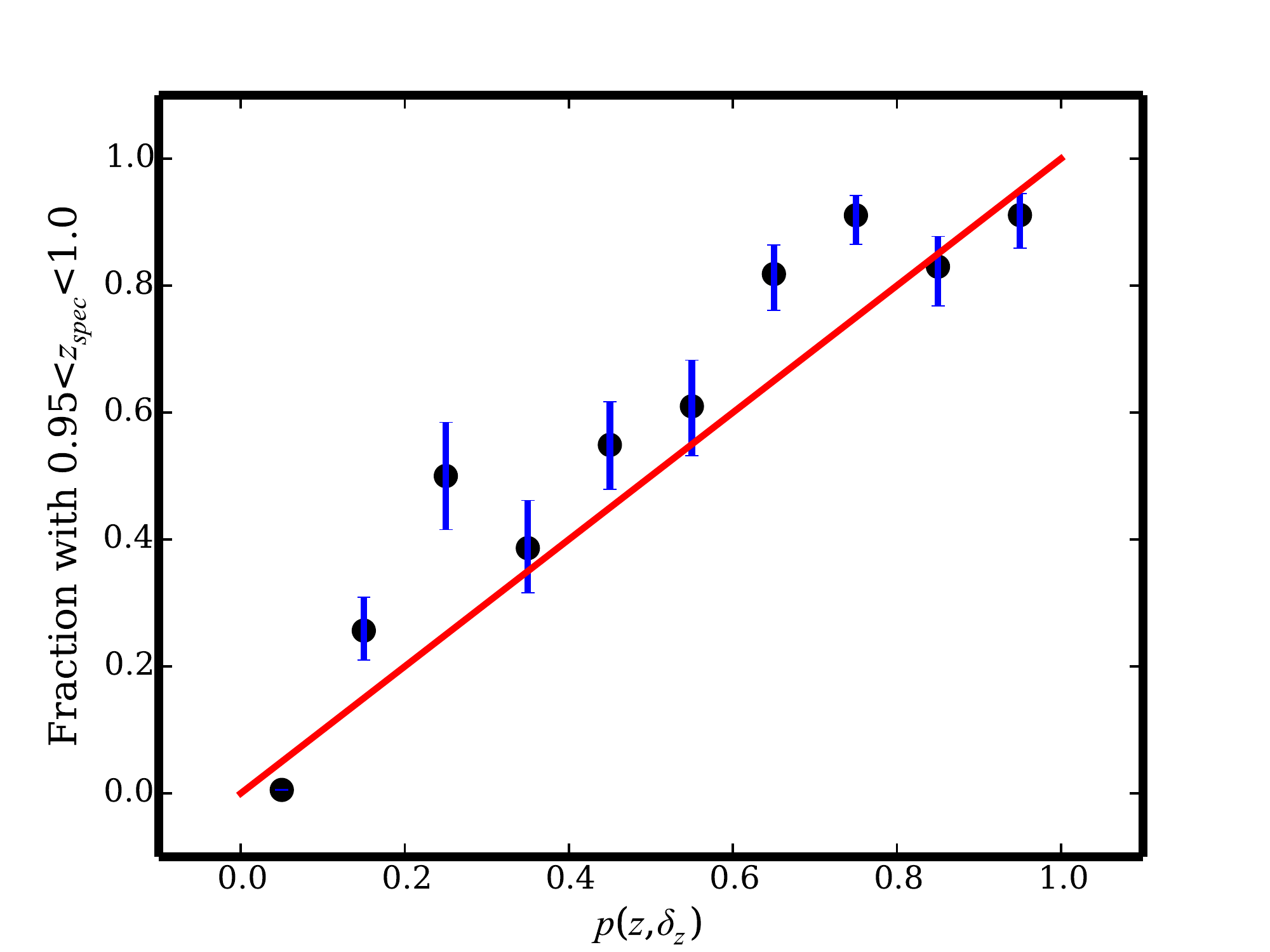}
\includegraphics[clip=true,trim=0mm 0mm 0mm 0mm,width=3in,angle=0]{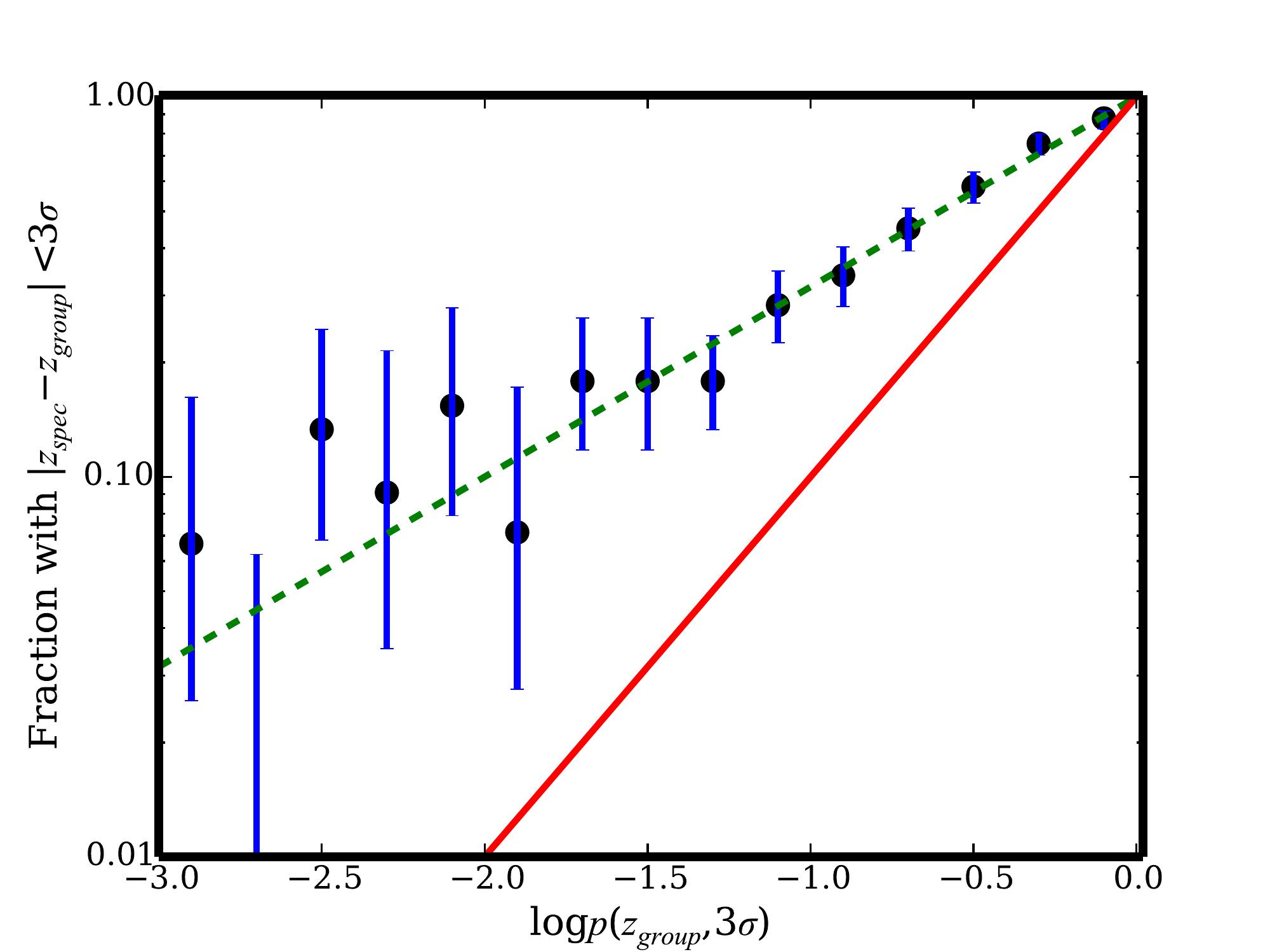}}
        \caption{{\it Left: }We show the fraction of galaxies with good quality spectra in our sample that lie in the redshift range $0.95<z<1.0$, as a function of the $p(z,\delta_z)$, the integral of the photometric redshift PDF within that range.  The solid line shows the one-to-one relation, for reference.  The two measures agree very well, indicating that $p(z,\delta_z)$ can be interpreted as the probability a galaxy actually lies at redshift $z$, on average.  {\it Right: } This shows the same quantities as in the left panel, but where $z=z_{\rm group}$ and $\delta_z=3\sigma$, for galaxies within 5.3 arcmin of each group centre.    The existence of an overdense structure along the line of sight means $p(z,\delta_z)$ underestimates the probability that a galaxy is a member of the group, as shown by the spectroscopic fraction on the y-axis.  We therefore use a simple fit to this relation, $\log{p_g}=0.5\log{p(z_{\rm group},3\sigma)}$, shown as the dashed line, and take $p_g$ as the probability that a galaxy is a member of a given group.}
\label{fig:fpz}
\end{figure*}

The statistical uncertainty on the velocity dispersions is determined using a jackknife technique, resampling from all candidate members and repeating the iterative method to determine membership and $\sigma$.  In addition, $\sigma$ is subject to bias due to the small number of members, and the clipping procedure.  We estimate this bias simply using Monte Carlo resamplings of a true Gaussian function, with the relevant number of members for each group.  In general the measured velocity dispersions underestimate the true dispersion, by up to $\sim 20$ per cent in the poorest systems.  Finally, we subtract in quadrature the typical rest-frame redshift uncertainty of $65$km/s, determined from repeat observations of galaxies (Paper~I).  This is most significant for the lowest-$\sigma$ systems, and thus partly offsets the bias noted above.  We refer to these corrected dispersions as the ``intrinsic'' velocity dispersion, $\sigma_i$, and base our physical analysis on them. There may be some concern that these dispersions are dominated by star--forming galaxies, which include more recently accreted galaxies and hence may be dynamically hotter than the passive population.  In general we have too few passive galaxies per cluster to measure a robust velocity dispersion from them exclusively.  For the three groups that have more than ten passive galaxies, we indeed find the velocity dispersion of the passive population to be smaller (70, 76 and 95 per cent of the adopted dispersion in groups 143, 71 and 120, respectively).  This is comparable to the level found in massive clusters at lower redshift \citep[e.g.][]{CNOC_dyn}. 

\subsection{Photometric Redshifts}\label{sec-photoz}
We make use of the photometric redshift catalogue of \citet[][v. 1.8]{Ilbert}.  For this work we are primarily interested in knowing which galaxies, lacking a spectroscopic redshift, are possible members of each group.   In general, integrating the PDF within a specific redshift range $z\pm\delta_z$ can be interpreted as the probability $p(z,\delta_z)$ that the galaxy lies in that range.  Both \citet{Ilbert} and \citet{George1} have shown this works fairly well.  We adopt a simple approach, where we model the redshift probability density function (PDF) as an asymmetric Gaussian, with $\sigma$ given by the 68th percentile on either side of the mean.  To demonstrate that this works for our specific case we show an example in Figure~\ref{fig:fpz}, where we plot the fraction of galaxies with good quality redshifts that lie in the redshift range $0.95<z<1.0$, in bins of $p(z, \delta_z)$. In this case, and for other relevant redshift ranges we have tried, the relation scatters about the 1:1 relation indicating that $p(z,\delta_z)$ provides a good description of the actual probability a galaxy is at redshift $z\pm \delta_z$.  

However, these distributions ignore the prior information that an overdense region exists in specific lines of sight, and therefore underestimates the true probability that a galaxy is within that overdense region.  To demonstrate this, in the right panel of Figure~\ref{fig:fpz} we repeat the above test within a few arcminutes of each group (corresponding to a GMOS field of view), and take $z=z_{\rm group}$ and $\delta_z=3\sigma$.  This is compared with the fraction of galaxies, $f$, in a given $p(z,\delta_z)$ bin, that are spectroscopically confirmed to lie within 3$\sigma$ of $z_{\rm group}$: this number is systematically larger.  The difference is largest at low probabilities, $p(z,\delta_z)<0.1$, where most of the galaxies are.  Therefore we consider the trend as a function of $\log{p(z,\delta_z)}$ and note that even for formally very low values ($\sim 10^{-3}$), approximately 10 per cent of targeted galaxies are group members.  We fit a linear relation $\log{p_g}=0.5\log{p(z_{\rm group},3\sigma)}$, and set $p_g=0$ for $p<0.001$.  We adopt this corrected value $p_g$ for the analysis.

In the remaining analysis of group populations, we include all galaxies in the COSMOS catalogue, weighted by this probability.  For galaxies with spectra, we assign a probability of unity if they are within the $3\sigma$ redshift limits of the group, and zero otherwise.   For calculating physical quantities like stellar mass and rest-frame colours, we use the median group redshift, rather than the peak photometric redshift probability.  Note that the same galaxy might be assigned to different groups, and in this case its estimated stellar mass will also be different, with $p_g$ capturing the probability that the redshift, and therefore the stellar mass, is correct. 
This approach has the advantage that, assuming the $p_g$ are correct, variations in completeness with spectral type, and from cluster to cluster, are accounted for.  It also allows 
 us to extend our analysis below the stellar mass completeness limit of $10^{10.3}$ adopted in Papers~I--III.  We caution, however, that results for these lower mass galaxies are sensitive to the correction made at low $p_g<0.1$, where most of the field galaxies are.  In practice the correction will depend, at least, on stellar mass, spectral type, and distance from the group centre.   We do not have a sufficient sample size here to make an empirical, reliable correction, and emphasize that robust results on this population requires spectroscopic follow-up.

\subsection{Group centres}

\begin{figure*}
\centerline{\includegraphics[clip=true,trim=0mm 0mm 0mm 0mm,width=3in,angle=0]{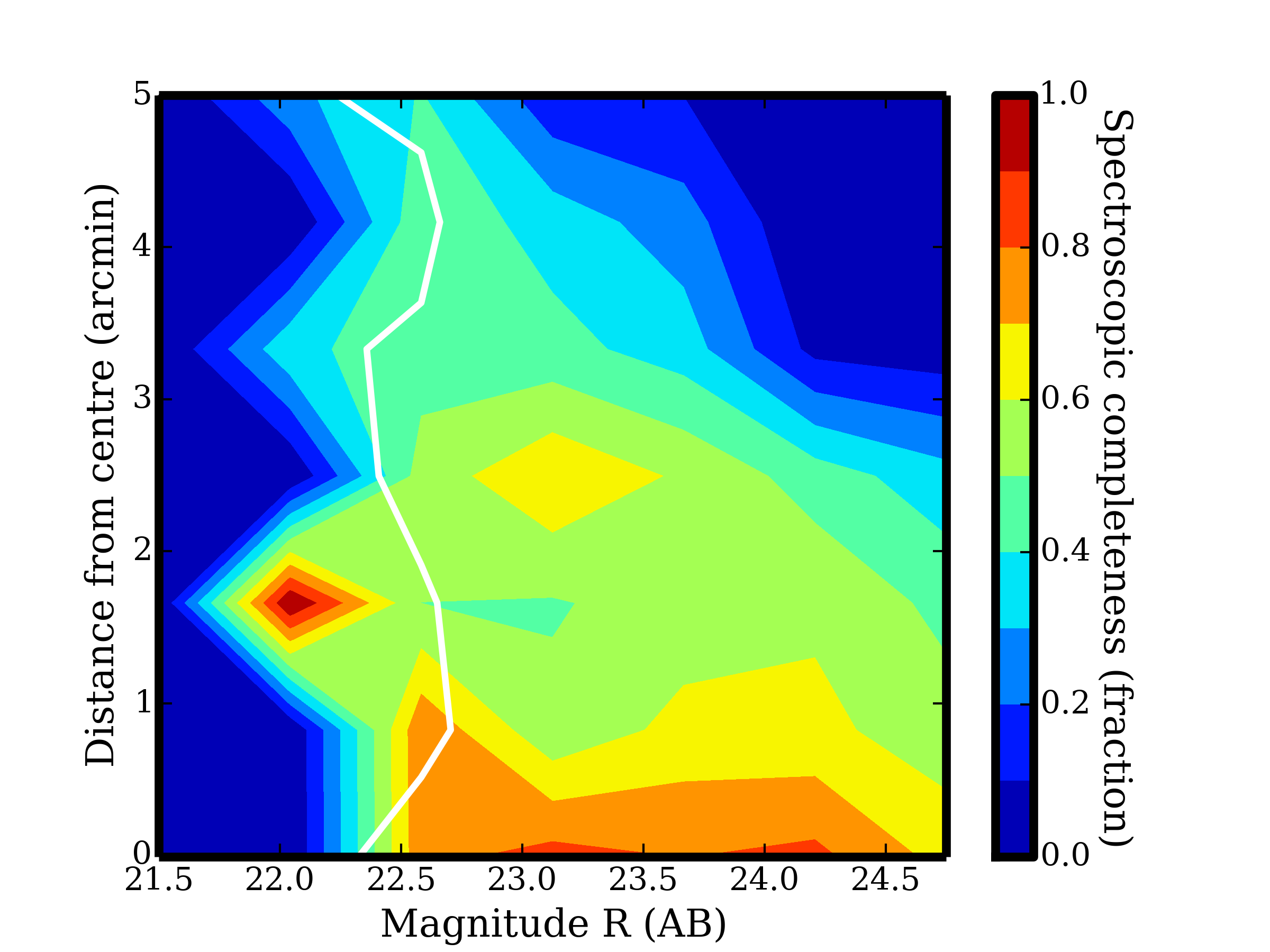}
\includegraphics[clip=true,trim=0mm 0mm 0mm 0mm,width=3in,angle=0]{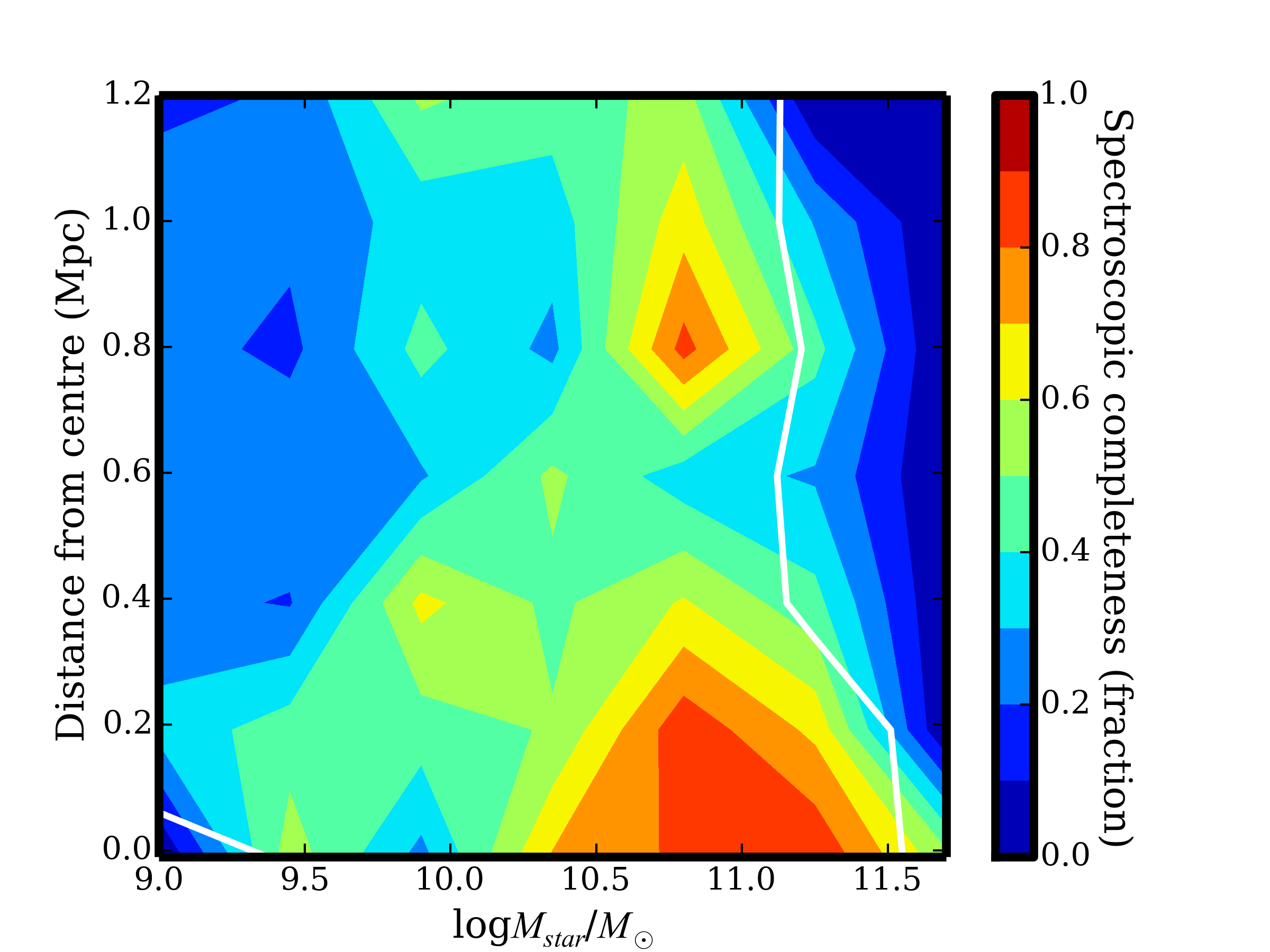}}
        \caption{{\it Left: }We show the spectroscopic completeness of group members in our sample, where the full group membership is estimated by integrating the $p_g$, suitably corrected for bias as described in Figure~\ref{fig:fpz}, of all galaxies.  This fraction is shown as a function of offset from the group centre in arcminutes, and r-band  magnitude.  Cells to the left of the white contour line (bright galaxies) have fewer than 5 photometric members each.  The spectroscopic completeness is $>50$ per cent for all galaxies within the GMOS field of view, brighter than our selection limit of $r=24.75$.  {\it Right: } The same completeness fraction is now shown as a function of physical parameters: spatial offset in Mpc, and stellar mass.  The sample is $>50$ per cent complete within the typical virial radius, for $M_{\rm star}>10^{10.3}M_\odot$.  Again the white contour indicates where the integrated $p_g$ per cell drops below 5 (to the right of the line in this representation).
\label{fig:comp}}
\end{figure*}
The initial estimate of the group centre comes from the updated X--ray catalogue of \citet{Fin07}, as used in \citet{George1}.  However, given the low X--ray flux in these systems we do not generally make use of this centre in further analysis.  Instead, we start with a centre defined by the average position of all spectroscopic members, and will refer to this as $C_{mean}$.  Membership is determined from the velocity distributions presented in \S~\ref{sec-velocities}, and some iteration is necessary to finalize a velocity dispersion, centre and radius that are all based on the same galaxies.

We will also consider an alterative centre, based on the stellar mass--weighted centroid of all passive galaxies.  This includes galaxies with photometric redshifts only, weighted by the appropriate value of $p_g$.  This centre will be referred to as $C_{q}$, and is generally close to $C_{mean}$ (see Table~\ref{tab-groups}).  For clarity in the analysis, however, we do not recompute the velocity dispersion, $R_{rms}$ or mean redshift based on this new centre.  

\subsection{Spectroscopic Completeness}
The spectroscopic completeness for group members is shown in Figure~\ref{fig:comp}.  The completeness is defined as the ratio of the number of spectroscopically-confirmed cluster members to the integrated $p_g$, and thus accounts both for the targeting strategy (including photo-z preselection) and redshift success rate.  In the left panel this is shown as a function of observable parameters: r-band magnitude and physical distance from the group centre in arcminutes.  The GEEC2 design was aimed at obtaining high completeness for $r<24.75$ within the GMOS field of view (field centre offsets of $\Delta r<2.75$ arcmin).  Figure~\ref{fig:comp} shows that this was achieved, with $>50$ per cent completeness in that range.  In the right panel we show the same quantity binned in physical parameters of stellar mass and clustercentric radius in Mpc.  This shows that the sample is $>50$ per cent complete within the typical virial radius, for $M_{star}>10^{10.3}M_\odot$.  These limits and completenesses are fully consistent with the values used in Papers~I--III, which were based on a simpler treatment of the photometric redshift uncertainties.  
\subsection{Stellar and Dynamical masses}\label{masses}

\begin{figure*}
\centerline{\includegraphics[clip=true,trim=0mm 0mm 0mm 0mm,width=3in,angle=0]{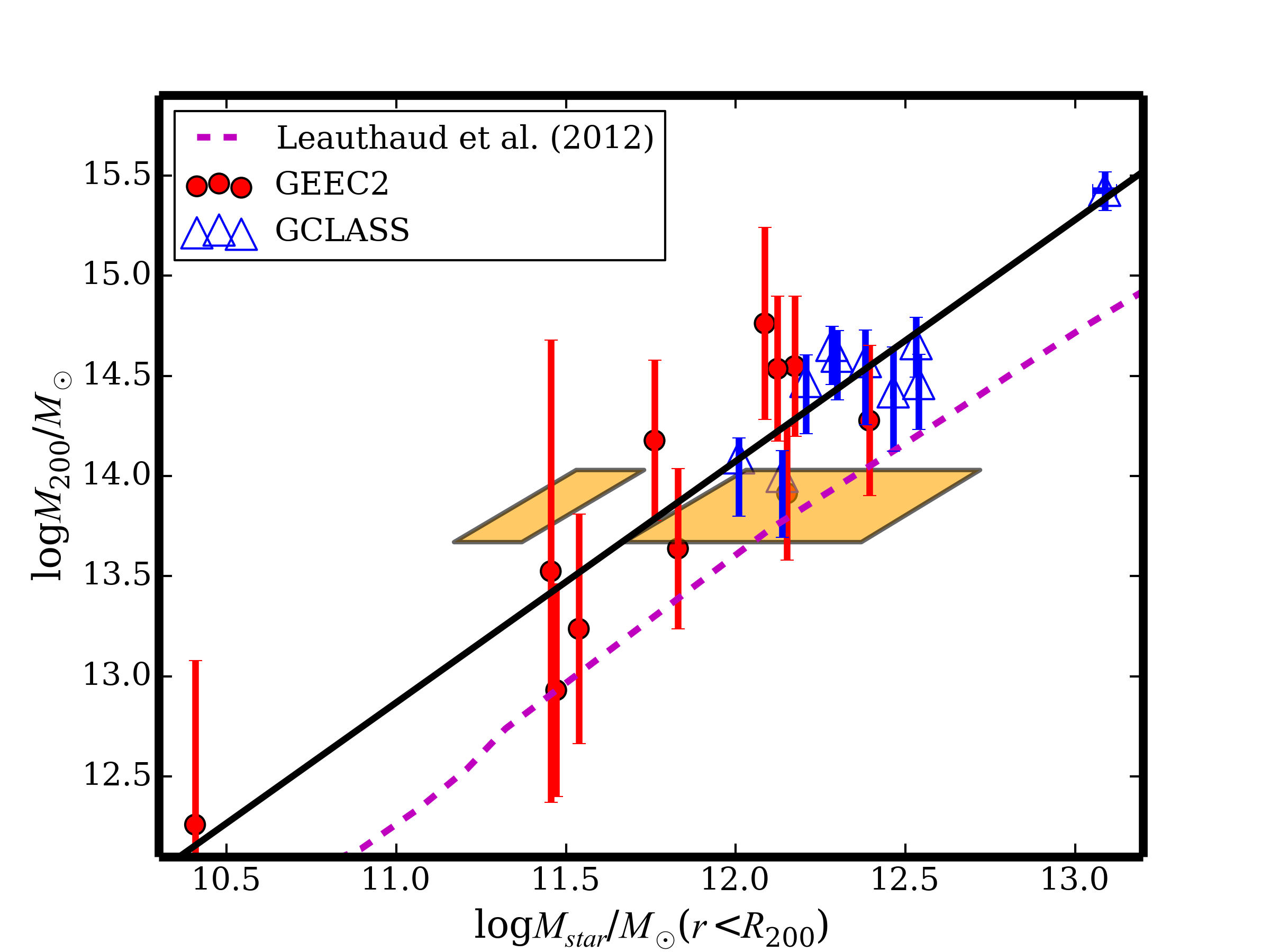}
\includegraphics[clip=true,trim=0mm 0mm 0mm 0mm,width=3in,angle=0]{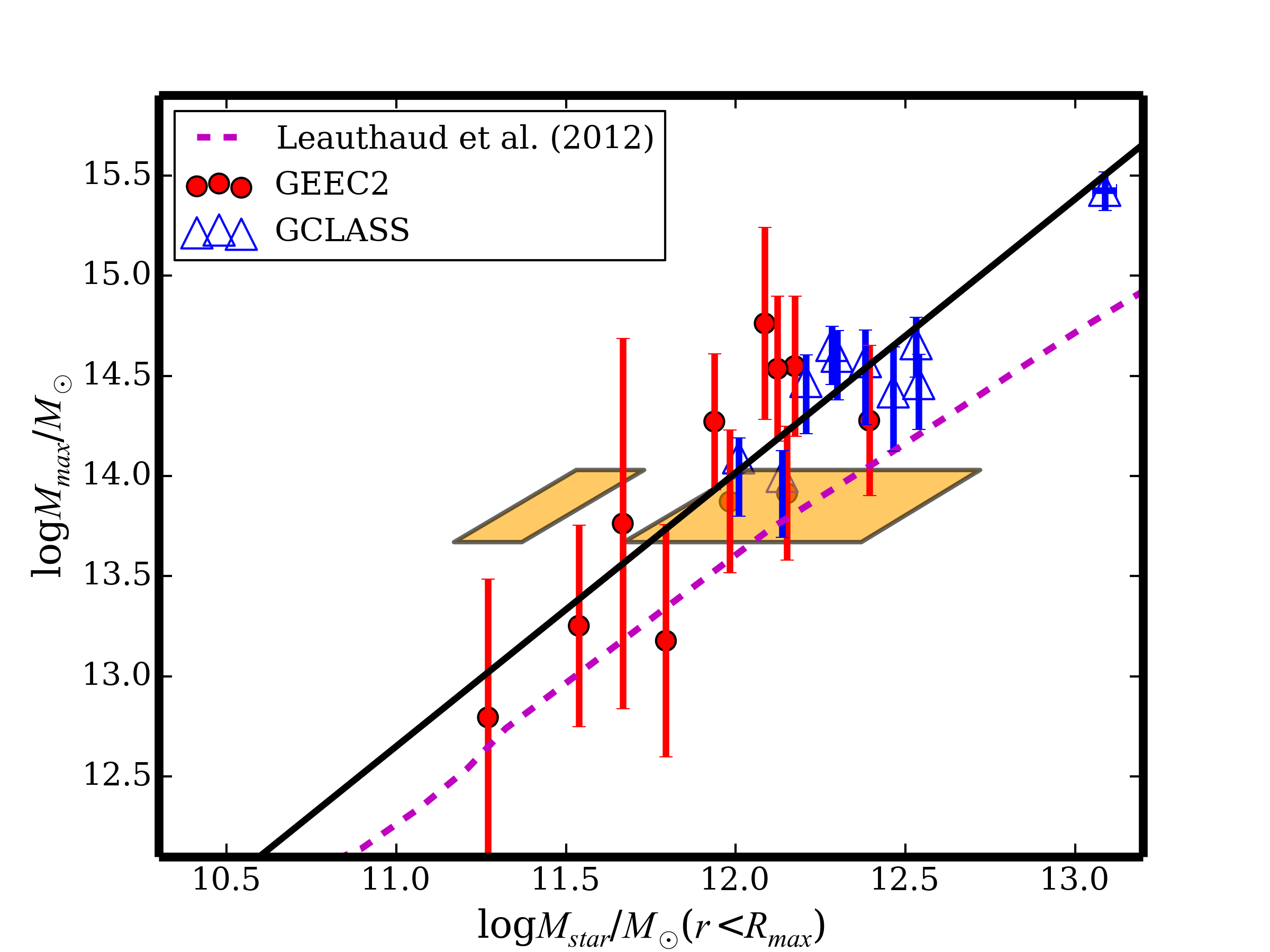}}
        \caption{{\it Left: } The correlation between total stellar mass within $R_{200}$ and dynamical mass, $M_{200}$, for the GEEC2 and GCLASS \citep{RvdB} clusters.  The solid line is the best fit, $\log{M_{200}}=1.20\left(\log{M_{\rm star}}-12\right)+14.07$.  The magenta line shows the HOD result from \citet{Leauthaud12}, while the orange boxes show the approximate range spanned by the X-ray groups as presented in \citet{Leauthaud-integrated}; the gap between them reflects a gap in the distribution of individual systems in that sample.  {\it Right:} The same, but for the total stellar mass within $R_{max}$ and $M_{max}$.  This only makes a difference for the lowest mass systems, for which $R_{rms}$ can be significantly larger than $R_{200}$.  Here the best fit is $\log{M_{max}}=1.36\left(\log{M_{\rm star}}-12\right)+14.02$.
\label{fig:MstarM200}}
\end{figure*}
We estimate the virial radius $R_{200}$, as the radius within which the average density is 200 times the critical density of the Universe at the redshift of the group.  By definition, assuming isotropic orbits and an isothermal distribution,
\begin{equation}
R_{200}=\frac{\sqrt{3}\sigma}{10H(z)},
\end{equation}
where $H(z)$ is the Hubble parameter at redshift $z$.  The corresponding dynamical mass is then
\begin{equation}
M_{200}=\frac{3R_{200}}{G}\sigma^2.
\end{equation}
We also calculate the quantity $M_{rms}$, where $R_{200}$ in the preceding equation is replaced with $R_{rms}$.  Recall that the velocity dispersion of passive galaxies may be smaller than the adopted values, by as much as 70 per cent, though this cannot be measured reliably for most groups in our sample.  If we assume the passive galaxies are better tracers of the potential, the dynamical masses of some groups may be overestimated by as much as a factor of two.  

The stellar mass of the group is computed as the total stellar mass of all galaxies, weighted by $p_g$ (recall $p_g=1$ for spectroscopically confirmed members).  The integration is carried out to $R_{200}$ and $R_{rms}$, resulting in two different estimates of this total. No explicit correction is made for galaxies below the mass limit of the photometric data, which reaches $M_{\rm star}<10^{9}M_\odot$.  In practice the spectroscopic completeness is high for $M_{\rm star}>10^{10}$ (see Figure~\ref{fig:comp}), and the contribution to the total stellar mass from from photometric members is generally less than 50 per cent.

For several of the groups with small $\sigma$, the corresponding $R_{200}$ is considerably smaller than the {\it rms} position, $R_{rms}$.  In other words, many of the galaxies for which $\sigma$ was computed lie outside $R_{200}$; in a couple of cases, depending on the choice of centre, there are {\it no} galaxies within $R_{200}$.  While one could consider whether this means the systems are unvirialized, or $\sigma$ is significantly underestimated, it is also possible that the assumption of a spherically--symmetric, isothermal sphere used to estimate $R_{200}$ can hardly be applied to individual systems with a handful of galaxies.
For these groups it might be more meaningful to consider the galaxies within $R_{rms}$ as members, and compute the mass within this radius instead.  Where relevant, then, we will also show the results for quantities computed within $R_{max}=\mbox{max}(R_{200},R_{rms})$, including the corresponding dynamical mass $M_{max}$.  In general this choice does not have a large impact on the correlations we show, apart from reducing some outliers corresponding to groups with very few members.  

In the left panel of Figure~\ref{fig:MstarM200} we show the correlation between the total stellar mass within $R_{200}$, and the dynamical mass $M_{200}$, for all groups in GEEC2.  We compare these with the published results from the ten more massive clusters in the GCLASS sample \citep{RvdB}, which lie in a similar redshift range.   The right panel shows the same, but for $R_{max}$ and $M_{max}$ instead.  The correlation is excellent, and the GEEC2 groups extend the relation defined by the higher mass GCLASS clusters.  On average, the stellar masses are 1\% of the dynamical masses.   While the statistical uncertainties on $M_{200}$ are large, they are formally negligible for the stellar masses.  The uncertainty on the latter is dominated by uncertainty in $R_{200}$, and while these are substantial, most of the stellar mass lies well within that radius.  

We compare these results with the halo occupation distribution (HOD) results at $z=1$ from \citet{Leauthaud12}, where we have converted their halo masses, measured relative to the background density, to ours using $M_{200,c}=0.87M_{200,b}$, appropriate for NFW haloes with concentration $c=2$ at $z=0.9$.  Our slope is almost identical to that of \citet{Leauthaud12}, with a normalization offset in the sense that our clusters have fewer stars for their dynamical mass than predicted.  In \citet{Leauthaud-integrated} this model was shown to be in good agreement with measurements of $\sim 20$ individual X--ray groups selected from COSMOS, drawn from the same catalogue followed up in the present work.  The range spanned by these data are shown with orange boxes on Figure~\ref{fig:MstarM200}, where we have converted $M_{500,c}=0.65M_{200,c}$.  We show two boxes to reflect the fact that there is a gap in the distribution of the actual data.  The median stellar mass of those groups is in perfect agreement with the HOD prediction, though the scatter is large. Some of the difference from our results might be attributable to velocity dispersions that are overestimated due to the dominance of low-mass emission line galaxies, but this is unlikely to be larger than a factor two (corresponding to a 30 per cent overestimate of $\sigma$), while the observed offset is $>2.5$.  Moreover we observe the same offset for the GCLASS clusters, which are much less prone to this problem.  

\begin{figure}
{\includegraphics[clip=true,trim=0mm 0mm 0mm 0mm,width=3in,angle=0]{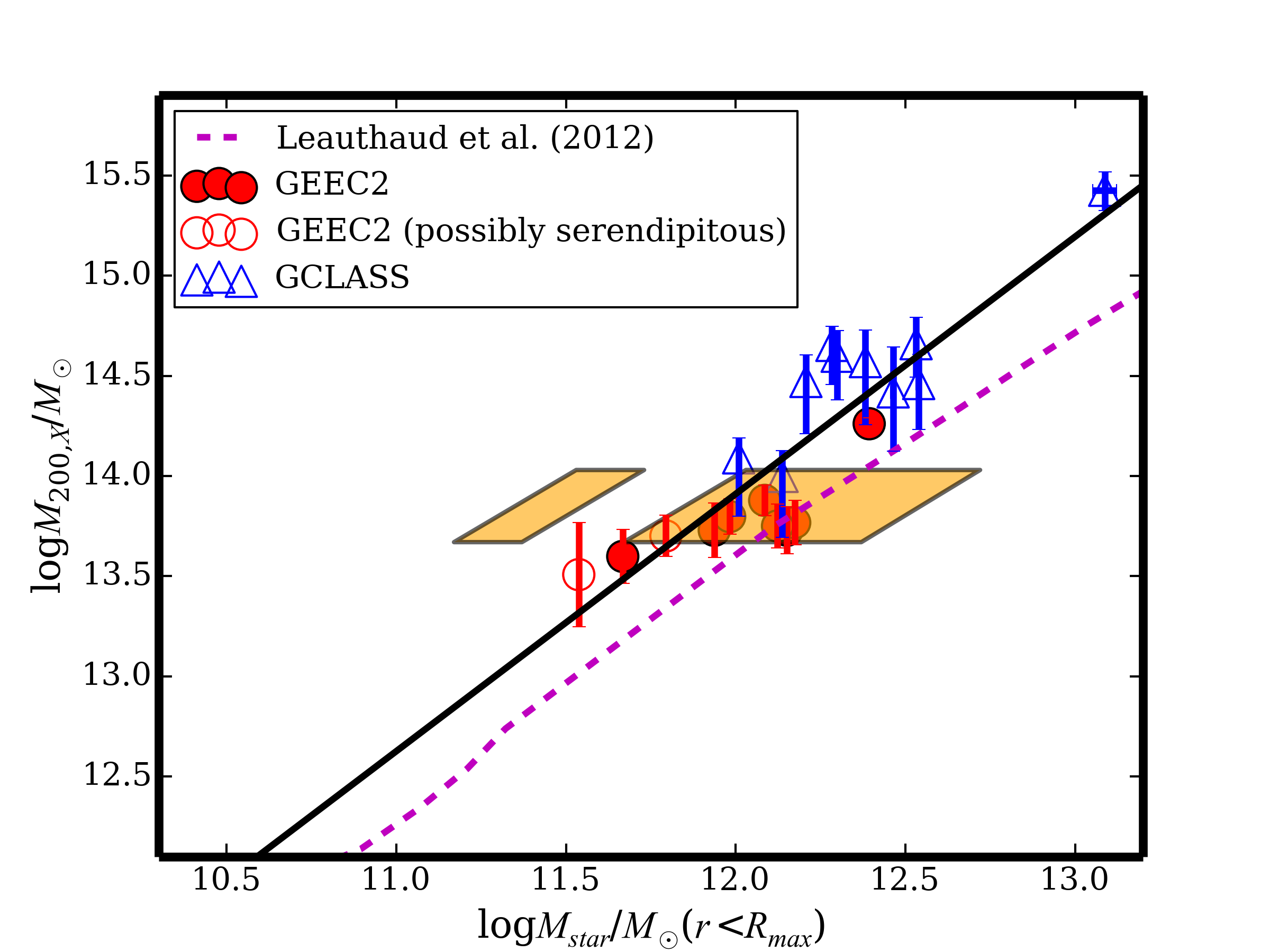}
}
        \caption{For the ten GEEC2 groups with X-ray associations, we show the correlation between total stellar mass within $R_{\rm max }$ and the mass estimated from the X-ray luminosity, assuming the calibration of \citet{COSMOS_lensing}.  The two groups (121 and 161) that may be unassociated with the observed X--ray emission are shown as open symbols.  The GCLASS clusters are shown for comparison, but here the dynamical mass estimate is still $M_{200}$ and is the same as in the previous figure. The best fit line, shown, is $\log{M_{200,x}}=1.28\left(\log{M_{\rm star}}-12\right)+13.90$.
\label{fig:MstarMx}}
\end{figure}

The GEEC2 groups were selected based in part on their X--ray luminosity, converted to a halo mass using the calibration of \citet{COSMOS_lensing}.  Group 213a, which was serendipitously found in the background of 213, is the only one not selected in this way, and it does not have an associated X-ray mass.  However, the redshift overdensities identified as groups 121 and 161, respectively, may not be associated with the targeted X-ray emission peak.  Thus we will generally treat these as serendipitous groups, as well. In Figure~\ref{fig:MstarMx} we show that the stellar mass within $R_{max}$ correlates well with this mass estimate, and still indicates a lower stellar fraction of $\sim 1$ per cent, compared with  \citet{Leauthaud12}.
The offset in stellar fraction observed in Figure~\ref{fig:MstarM200} is therefore unlikely to be related to a bias in the dynamical mass measurement.  
We note that a relatively low stellar fraction of $\sim 1$ per cent in groups at $z=1$ provides a natural link to the low stellar fractions in $z=0$ clusters, since it is difficult for this fraction to decrease with time \citep{B+08}.
\subsection{Summary of Group Properties}
\begin{table}
\begin{tabular}{lllllll}
Group & RA       & Dec          & $z$ &  $N_{\rm mask}$&$N_{\rm  mem}$ &$\sigma$\\ 
      &\multispan2{\hfil $C_{mean}$ (deg , J2000)\hfil} &           && &(km/s)    \\
\hline
40 & 150.414 & 1.848 & 0.9713 & 2 & 15 & $690\pm110$\\ 
71 & 150.369 & 1.999 & 0.8277 & 3 & 21 & $360\pm40$ \\
120 & 150.505 & 2.225 & 0.8361 & 3 & 32 & $550\pm60$\\
121 & 150.161 & 2.137 & 0.8374 & 3 & 5 & $80\pm40$ \\
130 & 150.024 & 2.203 & 0.9374 & 3 & 34 & $600\pm70$\\
134 & 149.65 & 2.209 & 0.9467 & 3 & 23 & $450\pm60$\\
143 & 150.215 & 2.28 & 0.881 & 3 & 19 & $580\pm70$ \\
150 & 149.983 & 2.317 & 0.9334 & 4 & 25 & $300\pm40$\\
161 & 149.953 & 2.342 & 0.944 & 4 & 8 & $170\pm30$\\
213 & 150.41 & 2.512 & 0.879 & 2 & 9 & $260\pm100$\\
213a & 150.428 & 2.505 & 0.9256 & 2 & 8 & $110\pm30$\\
\hline
\end{tabular}\caption{Properties of the eleven GEEC2 galaxy groups, as published in Paper~2.
Column 1 is the group ID number, and columns 2-6 give the position $C_{mean}$, median redshift, number of masks observed in each field, and number of group members, as indicated.   Column 7 is the rest-frame velocity dispersion as measured, without correction for any bias.  
\label{tab-groups_original}}
\end{table}

In this section we present a summary of the dynamical properties of each group.  In Table~\ref{tab-groups_original} we reproduce some of the properties as published in Paper~II.  These include the group centres $C_{mean}$, and the measured rest-frame velocity dispersion, uncorrected for biases noted above.  In Table~\ref{tab-groups} we present new derivations of dynamical quantities and mass estimates, based on the intrinsic $\sigma_i$ and the centre $C_q$.  This table also provides the stellar mass measurements within $R_{rms}$ and $R_{max}$.

\begin{table*}
\begin{tabular}{lllllllllll}
Group & RA      & Dec          &$\sigma_i$ &$R_{\rm rms}$&$M_{\rm rms}$&$M_{\rm star}/10^{11}M_\odot$&$R_{\rm 200}$&$M_{\rm 200}$&$M_{\rm star}/10^{11}M_\odot$&$M_{\rm 200,X}$\\
      &\multispan2{\hfil $C_q$ (deg, J2000)\hfil} &(km/s) &(Mpc)&($10^{13}M_\odot$)&$(r<R_{rms})$&(Mpc)&($10^{13}M_\odot$)& $(r<R_{200})$&($10^{13}M_\odot$)\\
\hline
40 & 150.422 & 1.850 & $830\pm 110$ & $0.34\pm 0.04$ & $16.2\pm 7.1$ & $6.0^{+6.2}_{-5.7}$ & $1.21\pm 0.19$ & $57.7\pm 27.6 $& $12.2^{+13.2}_{-11.4}$& $7.6\pm 0.6$\\
71 & 150.365 & 2.004 & $420\pm  40$ & $0.34\pm 0.03$ & $4.2\pm 1.3$ & $8.9^{+9.1}_{-8.3}$ & $0.67\pm 0.07$ & $8.2\pm 2.7 $& $14.2^{+14.8}_{-13.5}$& $5.4\pm 0.6$\\
120 & 150.507 & 2.224 & $560\pm  60$ & $0.81\pm 0.07$ & $17.4\pm 5.9$ & $22.2^{+24.7}_{-21.8}$ & $0.88\pm 0.11$ & $18.9\pm 7.1 $& $24.7^{+25.4}_{-21.9}$& $18.2\pm 0.5$\\
121 & 150.165 & 2.135 & $250\pm  40$ & $0.41\pm 0.05$ & $1.8\pm 0.9$ & $3.4^{+3.4}_{-3.4}$ & $0.40\pm 0.08$ & $1.7\pm 1.0 $& $3.4^{+3.4}_{-3.4}$& $3.2\pm 0.8$\\
130 & 150.025 & 2.207 & $700\pm  70$ & $0.71\pm 0.06$ & $24.1\pm 7.7$ & $10.0^{+11.8}_{-9.9}$ & $1.04\pm 0.12$ & $35.4\pm 12.4 $& $14.9^{+15.8}_{-14.2}$& $5.9\pm 0.6$\\
134 & 149.646 & 2.205 & $530\pm  60$ & $0.97\pm 0.07$ & $18.7\pm 6.3$ & $8.6^{+10.1}_{-7.7}$ & $0.78\pm 0.10$ & $15.0\pm 6.0 $& $5.8^{+7.5}_{-5.6}$& $5.4\pm 0.7$\\
143 & 150.211 & 2.280 & $680\pm  70$ & $0.23\pm 0.03$ & $7.5\pm 2.8$ & $8.6^{+8.7}_{-6.0}$ & $1.05\pm 0.13$ & $34.3\pm 12.4 $& $13.3^{+14.8}_{-12.4}$& $5.6\pm 0.6$\\
150 & 149.981 & 2.322 & $350\pm  40$ & $0.89\pm 0.08$ & $7.5\pm 2.7$ & $9.6^{+11.4}_{-9.3}$ & $0.52\pm 0.07$ & $4.3\pm 1.7 $& $6.8^{+7.7}_{-6.1}$& $6.3\pm 0.6$\\
161 & 149.974 & 2.341 & $200\pm  30$ & $0.53\pm 0.12$ & $1.5\pm 0.9$ & $6.2^{+7.3}_{-3.0}$ & $0.30\pm 0.05$ & $0.9\pm 0.5 $& $3.0^{+3.0}_{-3.0}$& $5.0\pm 0.5$\\
213 & 150.409 & 2.510 & $310\pm 100$ & $0.84\pm 0.13$ & $5.8\pm 5.3$ & $4.6^{+4.8}_{-4.0}$ & $0.49\pm 0.19$ & $3.3\pm 3.9 $& $2.9^{+3.1}_{-2.6}$& $4.0\pm 0.5$\\
9213 & 150.425 & 2.500 & $120\pm  30$ & $0.62\pm 0.09$ & $0.6\pm 0.4$ & $1.9^{+2.8}_{-1.7}$ & $0.18\pm 0.05$ & $0.2\pm 0.1 $& $0.3^{+0.3}_{-0.2}$& $-$\\

\hline
\end{tabular}\caption{Here we show the same groups as in Table~\ref{tab-groups_original}, but with the coordinates of the alternative centre, $C_{q}$ in columns 2-3.  Column 4 shows an estimate of the instrinsic velocity dispersion, where we have applied corrections for limited sampling, $2\sigma$ clipping, and redshift uncertainties.  This is combined with the $R_{rms}$ size (column 5, as published in Paper~2) to give an updated mass estimate $M_{rms}$ in column 6.  The stellar mass within this radius is given in column 6.  We also show $R_{200}$ and $M_{200}$, as determined from this intrinsic velocity dispersion, together with the stellar mass within this radius, in columns 8-10.  The final column shows the equivalent mass as estimated from the X--ray luminosity.  This is unavailable for the serendipitous group, 213a.  We note that groups 121 and 161 as we have defined them here may also be serendipitous, in the sense that they may not be associated with the targeted X-ray emission.
 \label{tab-groups}}
\end{table*}  

\begin{table}
\begin{tabular}{llllll}
Group & Galaxy & $z_{\rm spec}$ &  $z_{\rm phot}$&$p_g$&$\Delta R$ ($C_{q}$)\\
      &        & &&&(Mpc)\\
\hline

40 & 277060 & nan & $0.98^{+0.01}_{-0.02}$&0.75&1.84\\ 
40 & 277083 & nan & $1.04^{+0.03}_{-0.03}$&0.16&1.81\\ 
40 & 277335 & nan & $0.98^{+0.09}_{-0.03}$&0.50&1.86\\ 

...\\

213a & 1417695 & nan & $0.94^{+0.02}_{-0.02}$&0.27&0.87\\
213a & 1418640 & nan & $0.91^{+0.01}_{-0.01}$&0.18&1.18\\
213a & 1438646 & nan & $0.91^{+0.02}_{-0.05}$&0.25&1.09\\

\hline
\end{tabular}\caption{Sample entries from the online catalogue of group members.  We show the group and galaxy ID, the spectroscopic redshift (where available), the photometric redshift and its 1$\sigma$ uncertainty from \citet{Ilbert}, our bias--corrected measurement of $p_g$ and the offset from the mass-weighted centre of passive galaxies, $C_q$, in Mpc. The full table contains 1250 entries.}
\label{tab-grouptabs}
\end{table}
Our final sample consists of 162 spectroscopically--confirmed group members within $R_{\rm max}$.   The integrated $p_g$ within this radius, representing the effective group membership including those (mostly at low stellar masses) with photometric redshifts, is 400.
In Table~\ref{tab-grouptabs} we provide the members of each group with $M_{\rm star}>10^{9.5}M_\odot$ and within 2$R_{\rm max}$, including photometric members with $p_g>0.1$.  In addition to the galaxy ID we provide the spectroscopic redshift, where available, the photometric redshift from \citet[][v. 1.8]{Ilbert}, our bias--corected $p_g$, and the offset from $C_{q}$ in Mpc.

In the Appendix we show images of each group, with the radii $R_{200}$ and $R_{rms}$ shown as large white circles. The extent of the associated X--ray emission is illustrated with orange contours, and spectroscopic members are circled in red.  To provide a visual impression of the spectroscopic completeness, we identify photometric ``members'' with small white circles.  These members are identified probabalistically by drawing random numbers and deciding whether or not a given galaxy belongs to the group based on its $p_g$ probability.  Thus different galaxies will be identified in different realizations.

\section{The Most massive Cluster Galaxies}\label{sec:bcg}
We identify the most massive galaxies (MMG) in each group, including those without spectroscopic redshifts, within $R_{max}$ of the centre $C_q$.  In five groups there are two or more galaxies with similar stellar masses; particularly for galaxies which lack a spectroscopic redshift, there is some uncertainty about which is the most massive.  All but two of the alternative candidates are within $300$ kpc of the primary,  so proximity to the centre is also ambiguous, given centering uncertainties.   In Table~\ref{tab:bcg} we show the ID number of the identified MMG of each group, its stellar mass, $p_g$, and offset in Mpc from the mass-weighted centre of passive galaxies, $C_q$.  Where there is ambiguity as noted above, we show alternative choices.  The selected MMG is marked on the group images shown in the Appendix and is generally very close to the centre $C_q$.  Where useful, we will refer to the group centre as defined by the MMG as $C_{MMG}$.  Note that only one group, 130, has a MMG that is not spectroscopically confirmed.  In fact we do have a spectrum for that galaxy from zCOSMOS, and it has a redshift consistent with the group, but with a low confidence quality factor despite the fairly good quality spectrum (S/N$>5$ per pixel); it would likely have been class 3 or 4 in GEEC2.  All its physical properties are computed assuming it is at the mean redshift of the group.
\begin{table*}
\begin{tabular}{llllllllll}
Group&Galaxy& RA       & Dec           &z&$M_{\rm star}$&$p_g$&dR&$W_\circ$(OII)&SFR(SED)\\ 
     &      &\multispan2{\hfil (J2000 deg)\hfil} & &$(10^{11}M_\odot)$& &(Mpc)&(\AA)&$M_\odot/yr$\\
\hline
40 & 509928 & 150.42614 & 1.85492 & 0.9668 & 2.06 & 1.00 & 0.20 & $10.8\pm0.3$ & $0.52^{+0.03}_{-0.5}$\\
   & 510730 & 150.41315 & 1.85002 & - & 1.26 & 0.56 & 0.26 & - & \\
   & 510727 & 150.41386 & 1.84759 & - & 0.86 & 0.52 & 0.25 & - & \\
71 & 759679 & 150.36931 & 1.99949 & 0.8277 & 1.94 & 1.00 & 0.16 & $2.1\pm4.3$ & $0.64^{+0.19}_{-0.6}$\\
120 & 952745 & 150.50473 & 2.22425 & 0.8351 & 2.30 & 1.00 & 0.05 & - & $0.32^{+0.03}_{-0.31}$\\
   & 952744 & 150.50499 & 2.22508 & 0.8383 & 1.97 & 1.00 & 0.05 & $-1.5\pm14.5$ & $0.31^{+0.03}_{-0.29}$\\
   & 952743 & 150.50392 & 2.22444 & - & 1.63 & 0.78 & 0.07 & - & \\
   & 949127 & 150.50671 & 2.25378 & - & 2.55 & 0.77 & 0.83 & - & \\
121 & 1017862 & 150.16714 & 2.13210 & 0.8381 & 1.19 & 1.00 & 0.11 & $0.3\pm15.6$ & $0.5^{+0.2}_{-0.46}$\\
130 & 1032322 & 150.02382 & 2.20323 & - & 2.74 & 0.87 & 0.10 & - & \\
   & 1032173 & 150.02981 & 2.20744 & 0.933 & 1.09 & 1.00 & 0.15 & - & $0.38^{+0.05}_{-0.36}$\\
134 & 1081651 & 149.64963 & 2.20927 & 0.9535 & 1.33 & 1.00 & 0.18 & $16.0\pm3.8$ & $0.76^{+0.4}_{-0.72}$\\
143 & 995224 & 150.20776 & 2.28158 & 0.8821 & 1.47 & 1.00 & 0.10 & $16.3\pm5.0$ & $0.32^{+0.31}_{-0.31}$\\
150 & 1267155 & 149.98329 & 2.31718 & 0.9332 & 1.91 & 1.00 & 0.17 & - & $1.92^{+1.54}_{-1.8}$\\
161 & 1264300 & 149.99024 & 2.33671 & 0.9426 & 1.85 & 1.00 & 0.49 & $8.2\pm3.3$ & $0.27^{+0.13}_{-0.24}$\\
   & 1263957 & 149.97312 & 2.33802 & 0.9447 & 1.75 & 1.00 & 0.10 & $-0.4\pm10.9$ & $0.22^{+0.37}_{-0.2}$\\
   & 1262304 & 149.96143 & 2.34943 & - & 2.08 & 0.54 & 0.43 & - & \\
213 & 1411101 & 150.40967 & 2.51166 & 0.8785 & 1.61 & 1.00 & 0.06 & $17.9\pm11.1$ & $0.96^{+0.43}_{-0.9}$\\
9213 & 1410644 & 150.42643 & 2.51570 & 0.9259 & 0.49 & 1.00 & 0.44 & - & $3.65^{+1.52}_{-3.41}$\\
   & 1410661 & 150.42678 & 2.51018 & 0.9256 & 0.28 & 1.00 & 0.29 & $15.4\pm10.0$ & $0.08^{+0.08}_{-0.08}$\\
\hline
\end{tabular}
\caption{The properties of the most massive galaxies in each group.  Column (1) shows the group with which the galaxy is associated, and column (2) is the ID number of the galaxy itself.  Its position and spectroscopic redshift, where available, are given in columns 3-5.  The estimated stellar mass is given in column (6).  Column (7) shows $p_g$, which is unity for spectroscopically confirmed members.  The final column shows the proper distance, in Mpc, from $C_q$, the mass-weighted centre of passive galaxies.  For a few groups the choice is ambiguous and we list close alternatives as well.\label{tab:bcg}}
\end{table*}
\subsection{AGN}
We search for evidence of AGN activity in our MMG in the X--ray, mid-infrared, and radio data.  Strong, unresolved X--ray emission provides the most unambiguous measurement of AGN activity.  In order to distinguish the AGN emission from the diffuse, group-scale emission, we require the {\it Chandra} resolution, and thus use the point source catalogue from \citet{Elvis}.  Only one potential match is identified, at a distance of $\sim 1.3$ arcsec from the MMG in group 134.  While this source has a full band flux of $2.71\times 10^{-15}$erg~cm$^{-2}$~s$^{-1}$, it is undetected in the hard band and thus consistent with arising from the intergroup medium.  We also consider the IRAC colours, following \citet{Lacy04} and \citet{Stern05}, and find none of the MMGs in our sample show infrared excess expected of AGN--dominated emission.  We therefore have no convincing detections of X--ray AGN in our MMG sample. 

Deep 1.4 GHz radio data are also available from the VLA survey of \citet{Schinnerer07}.  Our MMG sample lie within the region of 20$\mu$Jy sensitivity.  To convert fluxes to rest-frame power, we use
\begin{equation}
P_{1.4 GHz}=4\pi D_\nu^2S_\nu,
\end{equation}
where $D_\nu=D_L(1+z)^{-(1+\alpha)/2}$ \citep{Birzan04}.  Following \citet{Hickox09} we assume a spectral index of $\alpha=0.5$.  Thus, the limits of our survey at $z=0.82$ and $z=0.97$ correspond to $\log{P_{1.4 GHz}}=22.69$ and $22.82$, respectively, well below the typical AGN power.  We find four MMGs have a match within 3 arcsec, as shown in Table~\ref{tab_radio}.  The radio emission in group 120 is spatially extended, and the power listed in Table~\ref{tab_radio} refers only to the central emission.  All but one of these shows power in excess of 23.8 used to distinguish AGN from star-forming galaxies in the study of \citet{Hickox09}, and that one, in group 150, has a power of $23.6\pm0.05$.  
\begin{table}
\begin{tabular}{lcc}
\hline 
Group & S$_{\mathrm{tot}}$ & Log$_{10}$ P$_{1.4 \mathrm{GHz}}$ \\ 
      & (mJy) & (W Hz$^{-1}$)  \\ 
\hline
         120 &     5.900     &    24.880 \\  
         134 &     0.551 $\pm$     0.032 &    23.946 $\pm$     0.025 \\ 
         143 &     1.088 $\pm$     0.028 &    24.186 $\pm$     0.011 \\ 
         150 &     0.243 $\pm$     0.028 &    23.579 $\pm$     0.047 \\
\hline
\end{tabular}
\caption{Radio fluxes and power in the MMG sample of GEEC2.}
\label{tab_radio}
\end{table}

\subsection{Morphologies and sizes}
In the appendix we present{\it HST} ACS F814W images of each MMG.  We use the reductions of \citet{Koekemoer07}, which included drizzling, flux calibration and registration.  The final images are sampled to 0.03$^{\arcsec}$ pixels and have a measured PSF FWHM of 0.09$^{\arcsec}$, or 0.7 kpc at $z=0.9$.  

The central galaxies of groups 121 and 213a are clearly disky, but the rest have early-type morphology.  It is interesting that groups 121 and 213a are unlikely to be associated with the observed X--ray emission, though the same may be true for group 161.  We measure 1D surface brightness profiles using the {\sc IRAF} {\sc ellipse} task, following \citet{Stott_r}, and cut out 500$\times$ 500 kpc regions around each MMG from the mosaic.  First we detect objects in the images using {\sc SExtractor} to confirm the apparent magnitude ({\sc MAG\_AUTO}) and to obtain approximate values for the central pixels.  We use initial estimates of the position and ellipticity based on single-S{\' e}rsic {\sc GIM2D} \citep{Gim2d} fits of \citet{Sargent07}.  We then use {\sc ellipse} in interactive mode, with the {\sc SExtractor} segmentation image as the basis of the contaminating object mask.  We extract the profile in bins of fixed physical width, to allow subsequent stacking of the data, and fit models of the form:
\begin{equation}\label{eqn-1dfit}
\mu(r)=\mu_e+c_n\left[\left(\frac{r}{r_e}\right)^{1/n}-1\right],
\end{equation}
where $\mu_e$ is the surface brightness at $r_e$, and $c_n=2.5b_n/\ln{10}$.  
\begin{table}
\begin{tabular}{ccccc}
\hline 
Group&\multispan2{\hfil $R_{\rm semi}$\hfil}& $n$&$b/a$\\
     &($n=4$, kpc)&(Free $n$, kpc)& &\\
\hline
          40 &   19.95 $\pm$    2.30 &  107.26 $\pm$  116.07 &   11.23 $\pm$   5.36 &    0.96 \\ 
          71 &    8.31 $\pm$    0.82 &    8.47 $\pm$    1.35 &    4.24 $\pm$    1.51 &    0.95 \\ 
          120 &   19.49 $\pm$    1.80 &  49.15 $\pm$    7.80 &   8.55 $\pm$    1.11 &    0.89 \\ 
         121 &    3.74 $\pm$    0.52 &    3.75 $\pm$    0.62 &    3.97 $\pm$    2.66 &    0.58 \\ 
         130 &    9.49 $\pm$    0.84 &   17.14 $\pm$   13.37 &   8.25 $\pm$    7.62 &    0.99 \\ 
         134 &   29.22 $\pm$    3.20 &  95.20 $\pm$    43.63 &    8.49 $\pm$    2.72 &    0.96 \\  
         143 &   13.15 $\pm$    1.60 &    8.54 $\pm$    0.82 &    1.91 $\pm$    0.42 &    1.00 \\  
         150 &   14.47 $\pm$    1.66 &   15.53 $\pm$    6.92 &    4.32 $\pm$    1.91 &    1.00 \\ 
         161 &    9.57 $\pm$    0.99 &   13.89 $\pm$    8.33 &    7.05 $\pm$    4.57 &    0.85 \\
         213 &   12.37 $\pm$    1.87 &    6.15 $\pm$    0.15 &    1.08 $\pm$    0.18 &    0.99 \\
         213a &   10.28 $\pm$    1.52 &    5.77 $\pm$    0.13 &    0.92 $\pm$  0.07 &    0.95 \\ 
\hline
\end{tabular}
\caption{One-dimensional surface brightness fit parameters to equation~\ref{eqn-1dfit} for the MMGs in our sample. \label{tab-1dfits}}
\end{table}

\begin{figure}
{\includegraphics[clip=true,trim=0mm 0mm 0mm 0mm,width=3in,angle=0]{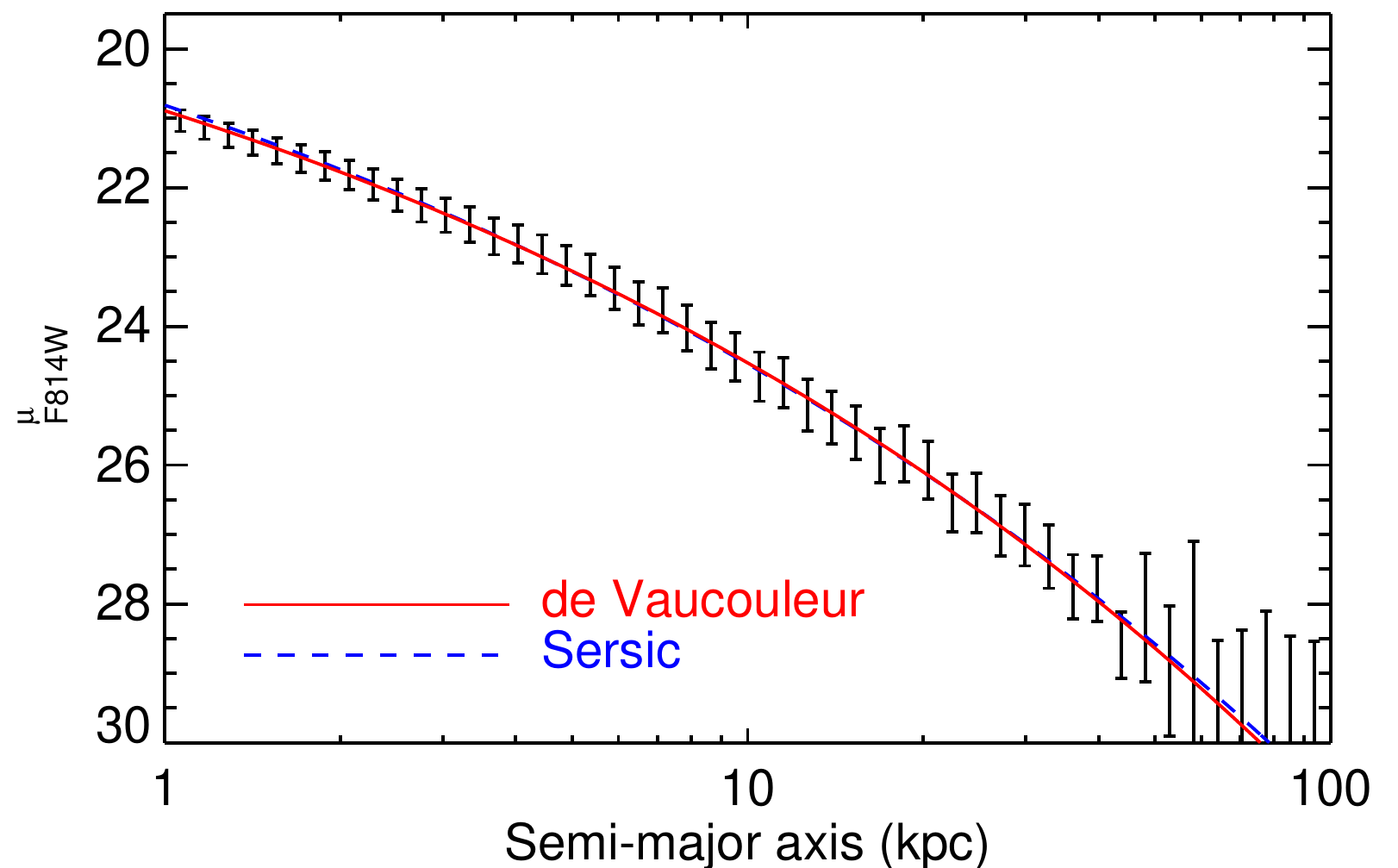}}
        \caption{The average surface brightness profile of the 11 MMG in our sample is shown, with the best-fit de Vaucouleurs ($R_e=10.1\pm 0.6$) and S{\'e}rsic ($R_e=10.3\pm 0.8$, $n=4.4\pm 0.8$) model fits.}
\label{fig:MMG_stack}
\end{figure}

In Table~\ref{tab-1dfits} we show the parameters for these fits to the MMG in our sample.   Results from both a fixed $n=4$ (de Vaucouleurs) model, and a free-n S{\' e}rsic model are shown.  In general, the axis ratios $b/a$ are high, reflecting the spheroidal appearance of most of the MMGs.  In Figure~\ref{fig:MMG_stack} we show the stacked profile of all 11 MMG and the corresponding fits.  The $n=4$ de Vaucouleurs model is an excellent fit to the average, with $R_e=10.1\pm 0.6$kpc.  Excluding the two disky galaxies from the fit results in a slightly larger effective radius of $R_e=11.0\pm 0.7$kpc, equivalent within the 1$\sigma$ uncertainties.

\begin{figure}
{\includegraphics[clip=true,trim=0mm 0mm 0mm 0mm,width=3in,angle=0]{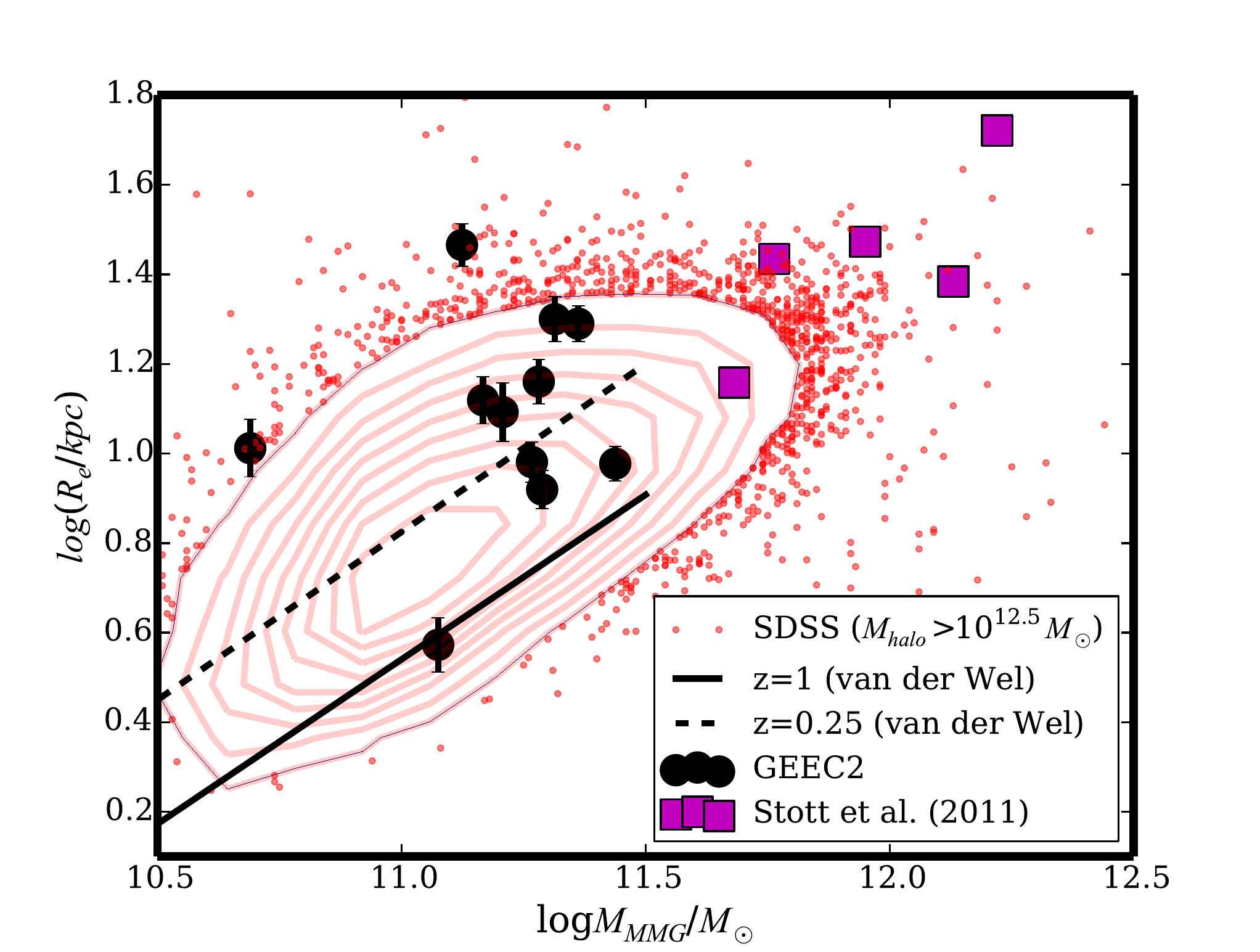}}
        \caption{The correlation between size and stellar mass of MMGs in our sample, compared with that of \citet{Stott_r} at similar redshift.  The solid and dashed line show the fit to passive galaxies in CANDELS, from \citet{CANDELS-MR}.  We also show the local relationship from SDSS as the red contours and points.  These are restricted to central, passive galaxies of haloes with $M>10^{12.5}M_\odot$ in the \citet{YangGC} catalogue, and the contours are weighted by the selection volume as calculated by \citet{Omand}.  The MMGs in our sample, and those of \citet{Stott_r}, have low stellar masses for their size, compared with the local sample.  This is consistent with substantial stellar mass growth, with limited growth in size. }
\label{fig-mass_size}
\end{figure}

We use the $n=4$ fits, for which the statistical uncertainties are smaller, to compare the correlation between stellar mass and $R_e$ with those in more massive clusters \citep{Stott_r,Stott_m}, as shown in Figure~\ref{fig-mass_size}.  
The GEEC2 MMG are both lower mass, and smaller in size, than those in the massive clusters.  They appear to lie approximately on an extension of the relation defined by those clusters, though the intrinsic scatter is large.  From recent analysis of CANDELS data, \citet{CANDELS-MR} have derived fits for the mass-size relation of passive and star--forming galaxies over $0.25<z<3$; we show their results for the passive population at $z=0.25$ and $z=1.0$ here.  The GEEC2 MMG are significantly larger for their mass than the $z=1$ relation; the same appears to be true for the \citet{Stott_r} MMG, though these are much more massive than anything in the CANDELS sample.  

We also include a comparison to local passive galaxies in the SDSS, using the $R_e$ measurements of \citet{Simard}.  We adopt the semi-major axis size from free-n Sersic fits, to be consistent with the analysis of \citet{CANDELS-MR}.  This local sample is restricted to central galaxies with $M>10^{12.5}M_\odot$ in the \citet{YangGC} catalogue, so the halo mass range approximately matches the range of our $z\sim 1$ sample.  The SDSS data are weighted to be representative of a volume-limited sample using the weightings published in \citet{Omand}. The peak of this distribution lies below the $z=0.25$ line of \citet{CANDELS-MR}, and we have confirmed that this is entirely due to the selection of massive haloes.  If we instead select all SDSS central galaxies, the agreement with \citet{CANDELS-MR} is excellent.  Thus it is apparent that the GEEC2 MMG are larger than even the typical $z=0$ central galaxy of the same mass.   The more natural interpretation of this is that the MMG in these haloes have grown more in stellar mass than they have in $R_e$.  Adopting the factor $1.8$ growth in stellar mass expected from the analysis of \citet{Lidman12}, with no evolution in $R_e$ brings most of the data into good agreement with the median $z=0$ relation.  The implication is that MMG at $z=1$ are already amongst the largest galaxies, with a size reflecting a unique formation process.  Further growth, perhaps through star formation, increases the mass while leaving the size relatively unchanged.  

\begin{figure*}
{\includegraphics[clip=true,trim=0mm 0mm 0mm 0mm,width=3in,angle=0]{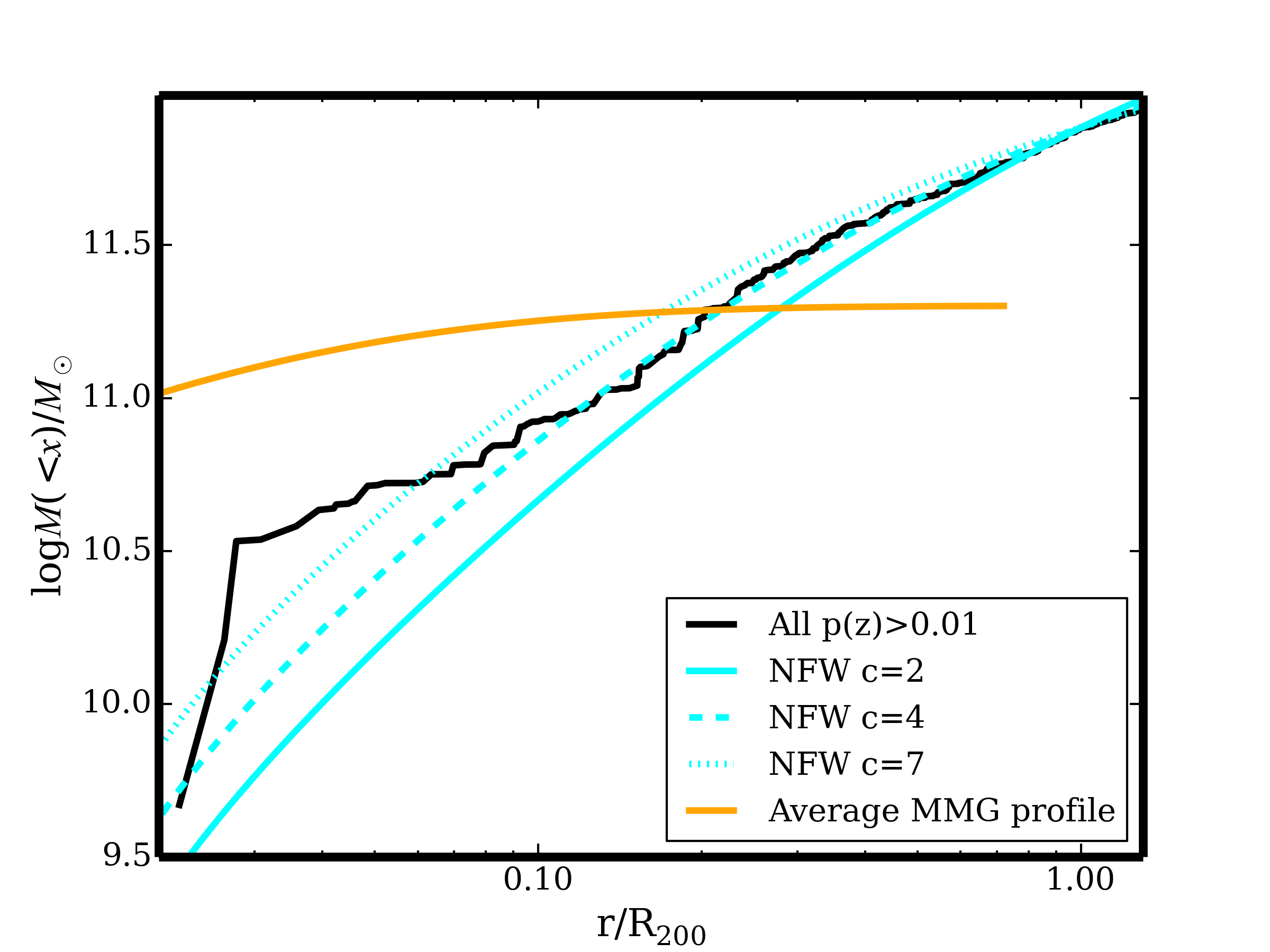}
\includegraphics[clip=true,trim=0mm 0mm 0mm 0mm,width=3in,angle=0]{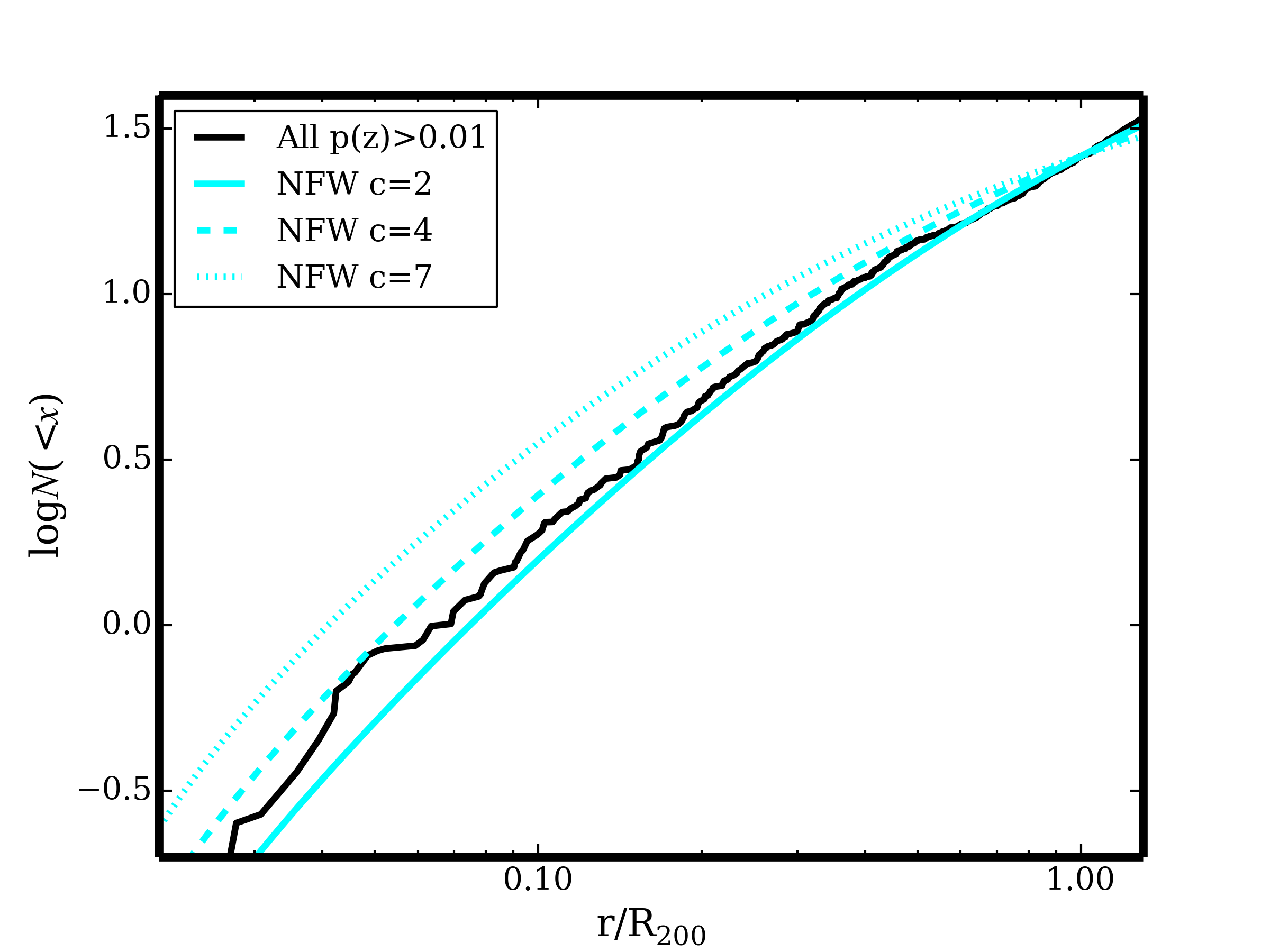}
}
        \caption{{\it Left: }The average cumulative stellar mass profile of galaxies in GEEC2 groups, including photometric members appropriately weighted by their probability of being in the group, but centred on, and excluding, the brightest group galaxy.  Three projected NFW mass distributions are plotted for comparison, normalized so the total mass within $R_{200}$ matches the total stellar mass of our groups within the same radius.  
The highest concentration model, $c=7$, is a good fit to the GCLASS results \citep{RvdB}, while the others ($c=2$ and $c=4$) are favoured by the dark matter simulations of \citet{Duffy} and \citet{DM14}, respectively. 
The orange line shows the average cumulative mass distribution of the MMGs in our sample, assuming $n=4$ with $R_e=13$kpc and $R_{200}=0.7$Mpc.  {\it Right: }The same, but for the enclosed number density of galaxies.   As found by \citet{RvdB}, the number density shows a lower concentration, indicating that the high concentration of stellar mass is due to galaxy mass segregation.}
\label{fig:mass_encl}
\end{figure*}
\section{Stellar mass distribution}
In Figure~\ref{fig:mass_encl} we show the cumulative distribution of stellar mass for the ensemble group, as a function of proper (physical) distance from the cluster.  For this we choose $C_{MMG}$ as the group centre, and exclude the MMG itself from the cumulative mass.  We do include galaxies without spectroscopic redshifts, weighted by $p_g$.  The location of the closest massive galaxy to $C_{MMG}$ is apparent from the steep increase in the cumulative function at small radius. 
Representative curves from three projected \citet[][NFW]{NFW} profiles are shown for comparison, normalized so the total mass within $R_{200}$ matches the average group stellar mass within this radius.  We confirm the GCLASS result of \citet{RvdB}, that the stellar mass distribution is centrally concentrated, and inconsistent with a concentration of $\sim 2$ that is seen in the dark matter simulations of \citet{Duffy}, for average groups at this redshift.  For $r>0.2R_{200}$ the fit is consistent with a $c=4$ profile,  in better agreement with recent simulation analysis of \citet{DM14}.  Within these central regions, however, there is an apparent excess of stellar mass, relative to the NFW profile.  This is consistent with our observations in Section~\ref{sec:bcg}, that the most massive galaxies in these groups have several nearby, massive companions (the steep increase at small $r$ reflects the location of the nearest MMG to its group).  Neither the high concentration, nor the central mass excess, is apparent in the number density profile (right panel), demonstrating that there is significant mass segregation in these groups, in contrast to the results of \citet{Ziparo13}.

The MMG effective radii range from $0.01$--$0.12$$R_{200}$, and thus the central galaxy light (which has not been included) would add to the central excess; the average contribution (for a $R_e=13$ kpc and $M=2\times 10^{11}M_\odot$ MMG) is illustrated with the orange curve in Figure~\ref{fig:mass_encl}.  With typical masses of $\sim 2\times10^{11}M_\odot$, the MMG mass distribution completely dominates in the region where the deviation from NFW is observed.  

\section{Discussion and conclusions}
GEEC2 is the first comprehensive, spectroscopic analysis of galaxies in group-mass haloes ($\sim 10^{13} M_\odot$) at $z\sim 1$.  With $162$ spectroscopically confirmed members among 11 groups we have been able to measure robust velocity dispersions, and hence halo masses, and to analyse the integrated stellar mass and its distribution.  With this paper we make available the redshift catalogue and most derived quantities.

We have combined the GEEC2 groups and GCLASS clusters to form a sample that spans over two orders of magnitude in halo mass.  The scaling relations between stellar and dynamical mass are fully consistent in the two samples.  From these we show that $\log{M_{200}}=1.20\left(\log{M_{\rm star}}-12\right)+14.07$, corresponding to a typical stellar fraction of 1 per cent, in good agreement with our $0.1<0.6$ measurement from groups in GEEC1 \citep{GEEC1_MstarMhalo}.  

Most of the groups in the sample have at least one massive galaxy near the centre, and we have measured the stellar light profile of the most massive galaxy from the {\it HST} images in the same way that \citet{Stott_r} have done for their more massive cluster sample.  We find that the MMG are well fit, on average, with a de Vaucouleurs profile with $R_e=10.1\pm 0.6$ kpc.  Our sample of 11 MMG are typically larger than the average galaxy at $z=1$ of similar mass, and even somewhat larger than the $z=0$ relation defined by haloes with $M>10^{12.5}M_\odot$.  This suggests the galaxies may grow more in mass than they do in size between $z=1$ and $z=0$.

Finally, we have shown that the stellar mass distribution in GEEC2 groups is well-represented by an NFW profile with a concentration of $c=4$.  This is lower than the $7\pm 1$ measured by \citet{RvdB} for GCLASS, but higher than measured in massive, lower redshift clusters \citep[][R. van der Burg et al., in prep.]{Muzzin_profiles}.  

\section{Acknowledgments}
\par
MLB would like to acknowledge generous support from NOVA and NWO visitor grants, as well as the hospitatility extended by the Sterrewacht of Leiden University, where this paper was written during sabbatical leave from Waterloo.
We are grateful to the COSMOS and zCOSMOS teams for making their excellent data products publicly available. This research is supported by NSERC Discovery grants to MLB and LCP. We thank the DEEP2 team, and Renbin Yan in particular for providing the {\sc zspec} software, and David Gilbank for helping us adapt this to our GMOS data. 

\bibliography{ms}

\clearpage
\appendix
\section{Group images and membership}
\subsection{Group images}
The appendix images are available in the published version from MNRAS.
Here we present Subaru $i$-band images centred on each group.  We indicate the position of spectroscopic members, with red circles.  Photometric members are selected at random, corresponding to their probability of being in the group $p_g$, and shown in white.  Thus, these images represent one realization, to give an impression of how many actual group members are likely missing from the spectroscopy.  Three large diamonds represent the three centres considered in this paper: blue for $C_{mean}$, white for $C_{MMG}$, and red for $C_{q}$.  We also indicate the two characteristic radii, $R_{rms}$ and $R_{200}$, as well as the extent of the X--ray emission from deep {\it Chandra} and {\it XMM} observations. 

The published appendix also includes HST ACS images of the MMG in each group.  

\end{document}